\documentclass{ifacconf}

\usepackage{graphicx}      
\usepackage{natbib}        
\usepackage{epsfig} 
\usepackage{mathptmx} 
\usepackage{times} 
\usepackage{amsmath}
\usepackage{subfig}
\usepackage{amssymb}
\usepackage{algorithm}
\usepackage{algpseudocode}
\usepackage{booktabs}
\usepackage{array}
\usepackage{textcomp}
\usepackage{url}
\usepackage{verbatim}
\usepackage{makecell}
\usepackage{caption}
\usepackage{xcolor}

\newtheorem{remark}{Remark}
\begin{document}
\begin{frontmatter}

\title{Algorithm Design and Comparative Test of \\ \textbf{N}atural Gr\textbf{a}dient Gaussia\textbf{n} Appr\textbf{o}ximation Filter} 

\thanks[footnoteinfo]{Wenhan Cao and Tianyi Zhang contribute equally to this paper. Corresponding Author: Shengbo Eben Li.}

\author[First]{Wenhan Cao} 
\author[First]{Tianyi Zhang} 
\author[First,Second]{Shengbo Eben Li}

\address[First]{School of Vehicle and Mobility, Tsinghua University, Beijing, China 
\\
(e-mail: \{cwh19, zhangtia24\}@ mails.tsinghua.edu.cn)
}
\address[Second]{College
of Artificial Intelligence, Tsinghua University, Beijing, China 
\\
(e-mail: lisb04@gmail.com)}

\begin{abstract}     Popular Bayes filters typically rely on linearization techniques such as Taylor series expansion and stochastic linear regression to  use the structure of standard Kalman filter. These techniques may introduce large estimation errors in nonlinear and non-Gaussian systems. This paper overviews a recent breakthrough in filtering algorithm design called \textit{N}atural Gr\textit{a}dient Gaussia\textit{n} Appr\textit{o}ximation (NANO) filter and compare its performance over a large class of nonlinear filters. The NANO filter interprets Bayesian filtering as solutions to two distinct optimization problems, which allows to define optimal Gaussian approximation and derive its corresponding extremum conditions.
The algorithm design still follows the two-step structure of Bayes filters.
In the prediction step, NANO filter calculates the first two moments of the prior distribution, and this process is equivalent to a moment-matching filter. In the update step, natural gradient descent is employed to directly minimize the objective of the update step, thereby avoiding errors caused by model linearization. Comparative tests are conducted on four classic systems, including the damped linear oscillator, sequence forecasting, modified growth model, and robot localization, under Gaussian, Laplace, and Beta noise to evaluate the NANO filter's capability in handling nonlinearity. Additionally, we validate the NANO filter's robustness to data outliers using a satellite attitude estimation example. It is observed that the NANO filter outperforms popular Kalman filters family such as extended Kalman filter (EKF), unscented Kalman filter (UKF), iterated extended Kalman filter (IEKF) and posterior linearization filter (PLF), while having similar computational burden.
\end{abstract}

\end{frontmatter}
\section{Introduction}
State estimation lies at the heart of numerous industrial control applications, including  robotics, power systems, manufacturing, aerospace, and transportation. In these domains, a crucial requirement is to effectively infer the internal state of a dynamical system from the information of noisy measurements and uncertain model. The measurements stem from various sensors and have noises with different distribution. The model has uncertain errors because one cannot describe the system behavior in a perfect manner. The key of filtering algorithm is to balance the uncertainties from noisy sensors and imperfect model.

Bayesian filtering emerges as a foundational framework for tackling state estimation problems by sequentially updating the probability distribution of the current state \citep{chen2003bayesian}.
When both the system dynamics and measurement models are linear with Gaussian noise, this framework simplifies to the well-known Kalman filter (KF) \citep{kalman1960new}. The KF’s elegant structure enables engineers to derive analytical expressions for the estimated mean and covariance, making it computationally attractive in linear systems.

\begin{figure*}
    \centering
    \includegraphics[width=0.65\linewidth]{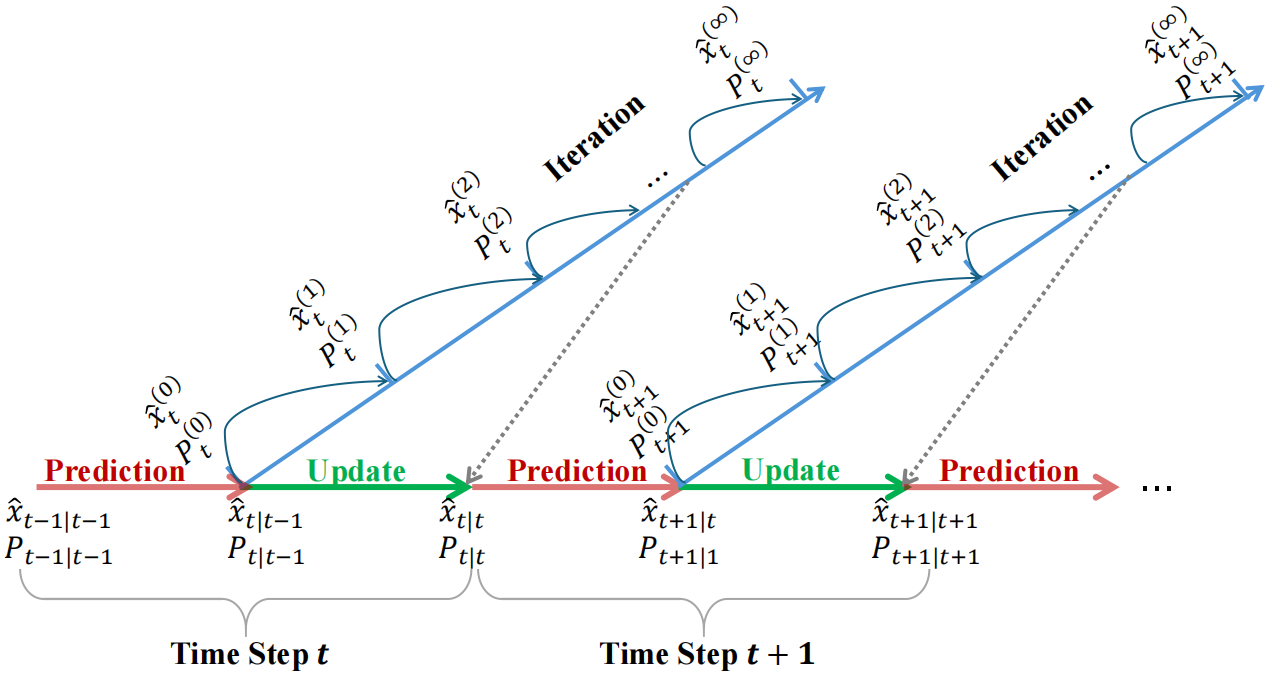}
    \caption{Illustration of NANO filter}
    \label{fig.NANO filter}
\end{figure*}

Nevertheless, the real world is often fraught with nonlinearities. In such scenarios, the standard KF struggles to deliver accurate state estimation due to its inherent linearity assumptions. Among existing solutions, particle filters leverage Monte Carlo sampling to approximate arbitrary probability distributions in nonlinear systems \citep{del1997nonlinear,liu1998sequential}. As another option, moving horizon estimation (MHE) directly computes maximum a posteriori estimates within a sliding time window \citep{cao2023generalized,cao2025robust}. However, both approaches suffer from prohibitive computational costs: particle filters require intensive sampling in high-dimensional spaces, while MHE involves solving nonlinear optimization problems at each time step. Such computational demands severely limit their applicability in time-sensitive industrial systems. As a computationally cheaper alternative, many practitioners turn to Gaussian filters, which restrict the state estimate to a Gaussian form at each time step, thus having
the potential to substantially reduce the computational burden \citep{thrun2002probabilistic}.

Gaussian filtering methods follow the philosophy of approximating the nonlinearities in the system and allows applying the well-established Kalman update mechanism. For example, the extended Kalman filter (EKF) and its iterated variant, namely the iterated extended Kalman filter (IEKF), linearize the system around an estimated operating point using Taylor expansions of the first order \citep{smith1962application,mcelhoe1966assessment,gelb1974applied}. Another approach is to employ stochastic linear regression, which leads to filters such as the unscented Kalman filter (UKF) \citep{julier1995new,julier2004unscented}, the Gaussian-Hermite Kalman filter \citep{arasaratnam2007discrete}, and the cubature Kalman filter \citep{arasaratnam2009cubature}. These methods aim to accurately capture the mean and covariance after a nonlinear transformation. Additionally, the posterior linearization filter (PLF) algorithm iteratively performs stochastic linear regression in the update step to further enhance estimation accuracy \citep{garcia2015posterior}.
Aforementioned Gaussian filters are often favored in real-time applications for nearly linear systems with Gaussian noise \citep{sarkka2023bayesian}. In practice, many industrial environments exhibit strong nonlinearities or complicated noise characteristics. Under these conditions, Gaussian filters may still introduce significant estimation errors. 

Motivated by these challenges, researchers continue to explore new Gaussian filtering approaches that retain the computational simplicity  while improving estimation accuracy to real-world systems. This paper outlines a tutorial on a recent breakthrough in the design of filtering algorithm, called \textit{N}atural Gr\textit{a}dient Gaussia\textit{n} Appr\textit{o}ximation (NANO) filter\citep{cao2024nonlinear}, which aims to enhance the performance of Gaussian filters under high nonlinearity. 
The NANO filter reconstructs Gaussian filtering as solutions to two parameter optimization problems, and derives its optimality conditions.
In the prediction step, NANO filter matches the first two moments of the prior distribution. In the update step, it directly minimizes the objective of the update step to avoid linearization errors in existing approachs. According to previous literature \cite{cao2024nonlinear}, NANO exists in two versions. One is derivative-based, which necessitates computing the first and second derivatives of the log-likelihood during iteration of the update step. The other is derivative-free, which leverages Stein's lemma to bypass these derivative calculations, leading to higher computational efficiency. This article will focus on the derivative-free version of NANO.

\section{Preliminaries}\label{sec.2}
Unscented transform (UT) is a computational tool for approximating the mean and covariance of a random variable after passing through a nonlinear function \citep{julier2004unscented}. Given a random variable \( x \in \mathbb{R}^n \) with mean \( \mu \) and covariance \( \Sigma \), UT is expressed as
\[
\{\mu', \Sigma'\} = \mathrm{UT}(\mu, \Sigma; f(\cdot)),
\]
where \( f(\cdot) \) is a nonlinear function. The core of UT is to compute the mean $\mu'$ and covariance $\Sigma'$ of \( y = f(x) \) with high accuracy, leveraging a set of sigma points generated from the distribution of \( x \). This process involves three steps:

\begin{enumerate}
    \item Sigma Point Generation: The \( 2n+1 \) sigma points $ \{\chi_0, \dots, \chi_{2n}\} $ are computed as 
    \[
    \chi_0 = \mu, \quad \chi_i = \mu \pm \sqrt{(n + \lambda) \Sigma}_i , \quad  i = 1, \dots, n.
    \]
    Here, \( \lambda \) is a scaling parameter, and \( \sqrt{(n + \lambda) \Sigma} \) is the Cholesky decomposition of \( (n + \lambda) \Sigma \).
    
    \item Propagation: The sigma points are propagated through the nonlinear function to obtain transformed points \( \chi'_i = f(\chi_i) \).
    
    \item Mean and Covariance Estimation:
    The new mean  and covariance are  weighted sums of the transformed points:
        \[
        \mu' = \sum_{i=0}^{2n} W_m^i \chi'_i, \quad
        \Sigma' = \sum_{i=0}^{2n} W_c^i (\chi'_i - \mu') (\chi'_i - \mu')^{\top},  
        \]
        with the weights being defined as 
        \begin{equation}\label{eq.UT}
        \begin{aligned}
                 W_m^0 &= \frac{\lambda}{n + \lambda}, \quad W_c^0 = W_m^0 + (1 - \alpha^2 + \beta),
                 \\
                 W_m^i &= W_c^i =  \frac{1}{2(n + \lambda)},
                 \quad
                 i = 1, \dots, 2n.
                 \end{aligned}    
        \end{equation}
    The parameters \( \alpha \), \( \beta \), and \( \lambda \) control the spread and weighting of the sigma points, with common defaults being \( \alpha = 0 \), \( \beta = 1 \), and \( \lambda = 0 \) for the so-called Julier's sigma points \citep{julier1995new}.
\end{enumerate}

\section{Algorithm Design}
\label{sec.NANO filter}
This section will overview the algorithm design of \textit{N}atural Gr\textit{a}dient Gaussia\textit{n} Appr\textit{o}ximation (NANO) filter, which was first proposed in the literature \citep{cao2024nonlinear}.  This filter reconstructs Gaussian filtering from an optimization perspective, where both the prediction and update steps are regraded as solutions to two optimization problems. The extreme condition for the problem in prediction step requires matching the first two moments of the prior distribution. In the update step, natural gradient that accounts for the curvature of the parameter space is derived to minimize the update step's objective to avoid linearization errors in Kalman filters family. 

Considering a nonlinear dynamic system with noisy measurements:
\begin{equation}\nonumber
\begin{aligned}
x_{t+1} &= f(x_t, u_t)+\xi_t,
\\
y_t &= g(x_t) + \zeta_t, 
\end{aligned}
\end{equation}
where $x_t \in \mathbb{R}^n$ is the system state, $u_t \in \mathbb{R}^l$ is the control input, $y_t \in \mathbb{R}^m$ is the noisy measurement. The function ${f}: \mathbb{R}^n \to \mathbb{R}^n$ is the transition function, and ${g}: \mathbb{R}^n \to \mathbb{R}^m$ is the measurement function; ${\xi}_t$ denotes a zero-mean  process noise with covariance matrix $Q \in \mathbb{R}_{+}^{n \times n}$, and ${\zeta}_t$ denotes a zero-mean measurement noise with covariance matrix $R \in \mathbb{R}_{++}^{m \times m}$. 

The procedure of NANO filter is illustrated in Fig. \ref{fig.NANO filter}. Its algorithm design still follows the two-step structure of Bayesian filtering: prediction and update, which means it also maintains a prior $p(x_t|y_{1:t-1}) $ and a posterior $p(x_t|y_{1:t}) $ at each time instant. As a Gaussian filter,
its key is to accurately approximate both the prior and posterior distributions as Gaussian distributions:
\begin{equation}\nonumber
\begin{aligned}
p(x_t|y_{1:t-1}) &\approx \mathcal{N}(\hat{x}_{t|t-1}, P_{t|t-1}), \\ p(x_t|y_{1:t}) &\approx \mathcal{N}(\hat{x}_{t|t}, P_{t|t}).   
\end{aligned}    
\end{equation}
Here, $\hat{x}_{t|t-1} \in \mathbb{R}^n$ and $ P_{t|t-1} \in \mathbb{R}^{n\times n}$ are called prior mean and prior covariance, and $\hat{x}_{t|t} \in \mathbb{R}^n$ and $ P_{t|t}\in \mathbb{R}^{n\times n}$ are called  posterior mean and poseterior covariance.
At each time instant $t$, NANO filter first performs a  prediction step and  then executes an update step. In the prediction step, UT is employed to calculate the prior mean and prior 
 covariance using the transition model. In the update step, the natural gradient descent technique is utilized to determine the optimal Gaussian posterior based on the measurement model and a new measurement. The NANO filter is summarized as follows:

\textbf{(1) Prediction Step}

(1-1) Calculate prior mean:
\begin{equation}\nonumber
\left\{ \hat{x}_{t|t-1}, W  \right\}= \mathrm{UT}(\hat{x}_{t-1|t-1}, P_{t-1|t-1}; f(\cdot, u_t)).
\end{equation}

(1-2) Calculate prior covariance:
\begin{equation}\nonumber
P_{t|t-1} = W+Q.    
\end{equation}

\textbf{(2) Update Step}

(2-1) Initialize posterior mean $\hat{x}_t^{(0)}$ and
posterior covariance $P_t^{(0)}$ at each time instant $t$. Here, $P_t^{(0)}$ is required to be positive definite.  

(2-2) For $k$, do iteration:

\quad \quad (i) Define three new functions:
\begin{equation}\nonumber
\begin{aligned}
L(z) &= \frac{1}{2} \|y_t - g(z) \|^2,
\\
L_x(z) &= \frac{1}{2} \left(z - 
\hat{x}_t^{(k)}
\right) \|y_t - g(z) \|^2,
\\
L_{xx}(z) &= \frac{1}{2} \left(z - 
\hat{x}_t^{(k)}
\right) \left(z - 
\hat{x}_t^{(k)}
\right)^{\top} \|y_t - g(z) \|^2.
\end{aligned}    
\end{equation}

\quad \quad (ii) Calculate expected coefficients:
\begin{equation}\nonumber
\begin{aligned}
\left\{ V, \cdot  \right\} &= \mathrm{UT}(\hat{x}_{t}^{(k)}, P_{t}^{k}; L(\cdot)),
\\
\left\{ V_x, \cdot  \right\} &= \mathrm{UT}(\hat{x}_{t}^{(k)}, P_{t}^{k}; L_{x}(\cdot)),
\\
\left\{ V_{xx}, \cdot  \right\}&= \mathrm{UT}(\hat{x}_{t}^{(k)}, P_{t}^{k}; L_{xx}(\cdot)).
\end{aligned}
\end{equation}

\quad \quad (iii) Calculate intermediate mean and covariance:
\begin{equation}
\label{eq.inter_mean}
\begin{aligned}
S^{(k)} =& \left(P_t^{(k)}\right)^{-1},
\\
S^{(k+1)} =& P_{t|t-1}^{-1} 
+ \alpha S^{(k)}(
V_{xx} S^{(k)} - V),
\\
P_{t}^{(k+1)} = &(S^{(k+1)})^{-1},
\\
\hat{x}^{(k+1)}_t =& \hat{x}^{(k)}_t - \alpha P_t^{(k+1)}( S^{(k)}
V_{x} + P_{t|t-1}^{-1}(\hat{x}_t^{(k)} - \hat{x}_{t|t-1})).
\end{aligned}    
\end{equation}

\quad \quad (iv) Stopping criterion: 
\begin{equation}\label{eq.stopping criterion}
\|P^{(k+1)}_t - P^{(k)}_t \|
< \gamma.
\end{equation}

(2-3) Update posterior mean and posterior covariance
\begin{equation}\nonumber
\hat{x}_{t|t} = \hat{x}_t^{(k+1)}, \quad
P_{t|t} = P_t^{(k+1)}. 
\end{equation}

\begin{remark}
In step (2-1), one can use the prior mean and prior covariance to initialize posterior parameters. Sometimes, this may lead to unstable iteration. A better initialization should facilitate the convergence of the update step and help guarantee the positive definite property of intermediate covariance matrices. As suggested by \citep{cao2024nonlinear}, we can use EKF to do posterior initialization. In this case, \(P_t^{(0)}\) is initialized as  
\begin{equation}\nonumber
\begin{aligned}
P_t^{(0)} =& P_{t|t-1}^{-1} + \left[ g'(\hat{x}_{t|t-1}) \right]^\top R^{-1} g'(\hat{x}_{t|t-1}) \\
&-  \sum_{j=1}^m \left[ y_{t} - g(\hat{x}_{t|t-1}) \right]_j g_j''(\hat{x}_{t|t-1}), 
\end{aligned}  
\end{equation}
where \(g'(\hat{x}_{t|t-1})\) is the Jacobian matrix of \(g(\cdot)\) with respect to \(\hat{x}_{t|t-1}\), and \(g''_j(\hat{x}_{t|t-1})\) is the Hessian matrix of the \(j\)-th component of \(g(\cdot)\) with respect to \(\hat{x}_{t|t-1}\).
\end{remark}

\begin{remark}
The stopping criterion in \eqref{eq.stopping criterion} uses the \(l_2\) norm of the error in two consecutive covariances. A more practical criterion is to avoid premature termination
of algorithm iteration. As suggested by \citep{cao2024nonlinear}, one can use the KL divergence of  Gaussian distributions of consecutive steps as the stopping criterion:
\begin{equation}\nonumber
D_{\mathrm{KL}}\left(\mathcal{N}\left(\hat{x}_t^{(k)}, P_t^{(k)} \right) \| \mathcal{N}\left(\hat{x}_t^{(k+1)}, P_t^{(k+1)} \right)\right) < \gamma,
\end{equation} 
where $\gamma>0$ is a predefined threshold.
\end{remark}

\begin{remark}
UT is not the only option in the design of NANO filter.
The purpose of UT is to calculate the mean and covariance of a random variable after it passes through a nonlinear function. Alternative methods such as Gaussian-Hermite approximation \citep{arasaratnam2007discrete} or spherical cubature integration \citep{arasaratnam2009cubature} can be used instead. Which one is chosen depends primarily on computational efficiency and computational accuracy, with UT being widely used as it provides a good balance between efficiency and accuracy. 
\end{remark}

\begin{remark}
Julier's sigma points are not the only choice for UT. More advanced sampling rules, such as Van der Merwe's sigma points \citep{van2003sigma} (i.e., with $\alpha = 10^{-3}$, $\beta = 1$ and $\lambda = 3-n$ in \eqref{eq.UT}), are also applicable. For NANO filter, it is observed that using Van der Merwe's sigma points in the update step can lead to unstable gradient calculations. A practical compromise is to use Van der Merwe's sigma points in the prediction step and Julier's sigma points in the update step.
\end{remark}

\begin{remark}
The step size $\alpha$ in \eqref{eq.inter_mean} is a tunable parameter in the update step. 
Theoretical analysis in~\citep{cao2024nonlinear} has proved that when $\alpha=1$, NANO filter is locally converges to the optimal Gaussian approximation at each time step, with accuracy up to a second-order remainder in the Taylor expansion. The estimation error is proven exponentially bounded for nearly linear measurement equation and low noise levels through contructing a supermartingale-like property across consecutive time steps. In practice, a parameter value between $0$ and $1$ is also acceptable in NANO filter, and sometimes it may results in smaller estimation error.
\end{remark}

\section{Experimental Results}\label{sec.experiments}
\begin{figure}[!t]
\centering
\subfloat[Case A: Gaussian noise]{
    \includegraphics[width=0.45\textwidth]{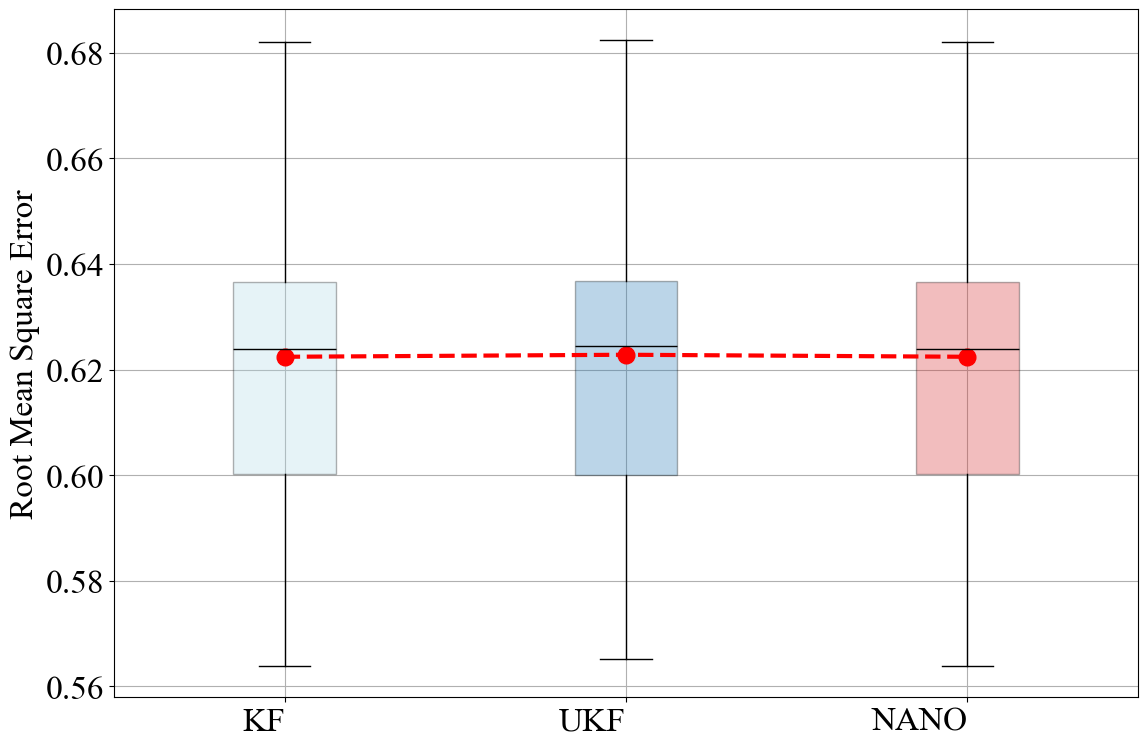}
    \label{fig.osci_gauss}
}
\\
\subfloat[Case B: Laplace noise]{
    \includegraphics[width=0.45\textwidth]{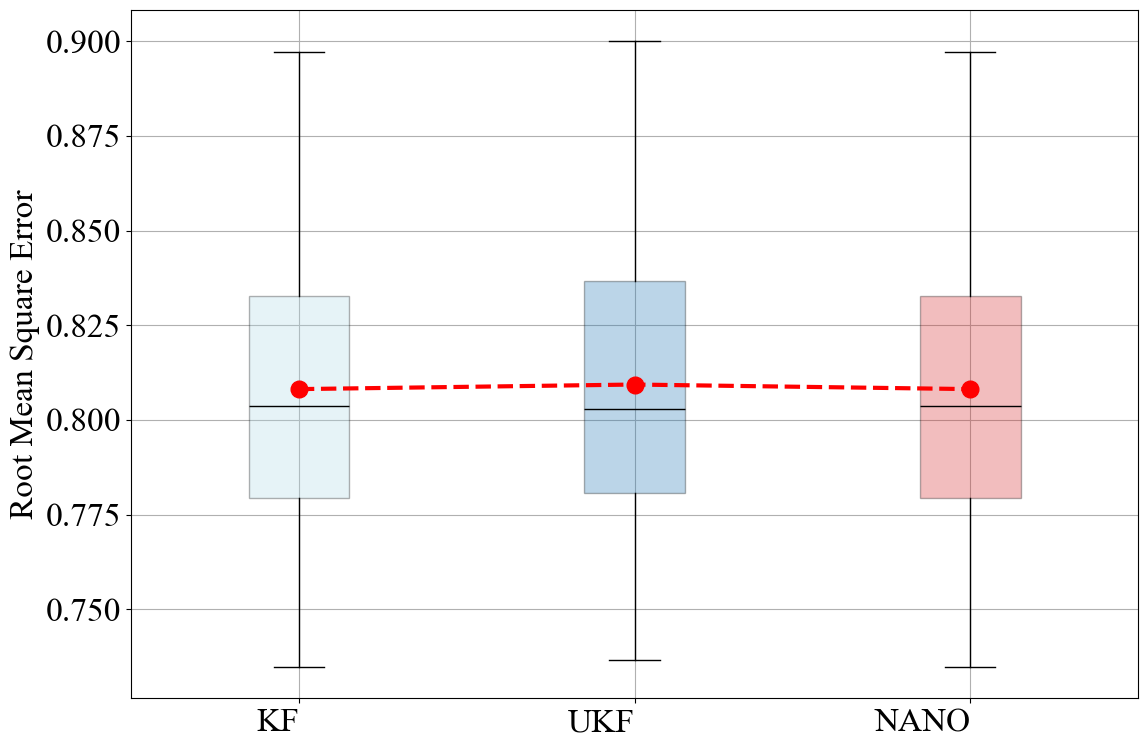}
    \label{fig.osci_la}
}
\\
\subfloat[Case C: Beta noise]{
\includegraphics[width=0.45\textwidth]{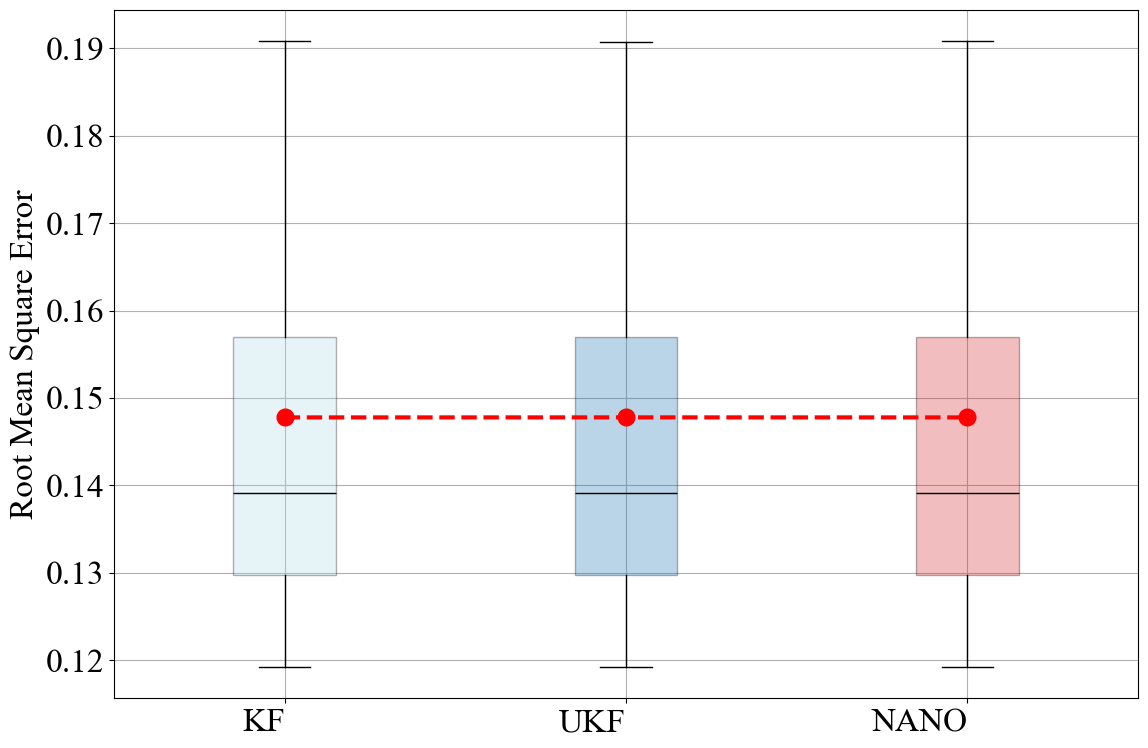}
\label{fig.osci_beta}
}
\caption{Box plot of RMSE over all MC experiments for damped linear oscillator. Note that red point `` $\textcolor{red}{\bullet}$
 " represents the average RMSE.}
\label{fig.osci}
\end{figure}

\begin{figure}[!t]
    \centering
    \subfloat[Case A: Gaussian noise]{
    \includegraphics[width=0.225\textwidth]{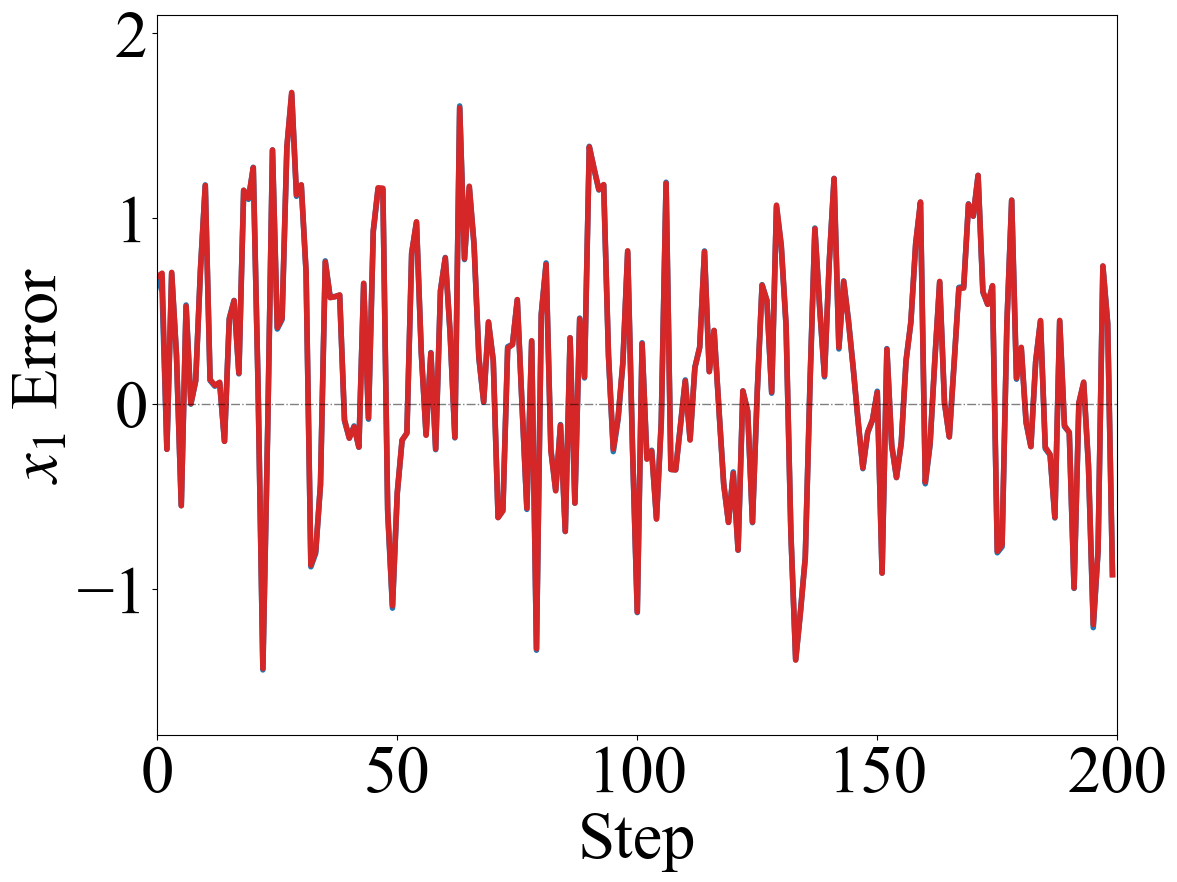}
    \label{osci:1}
    \hfill
    \includegraphics[width=0.225\textwidth]{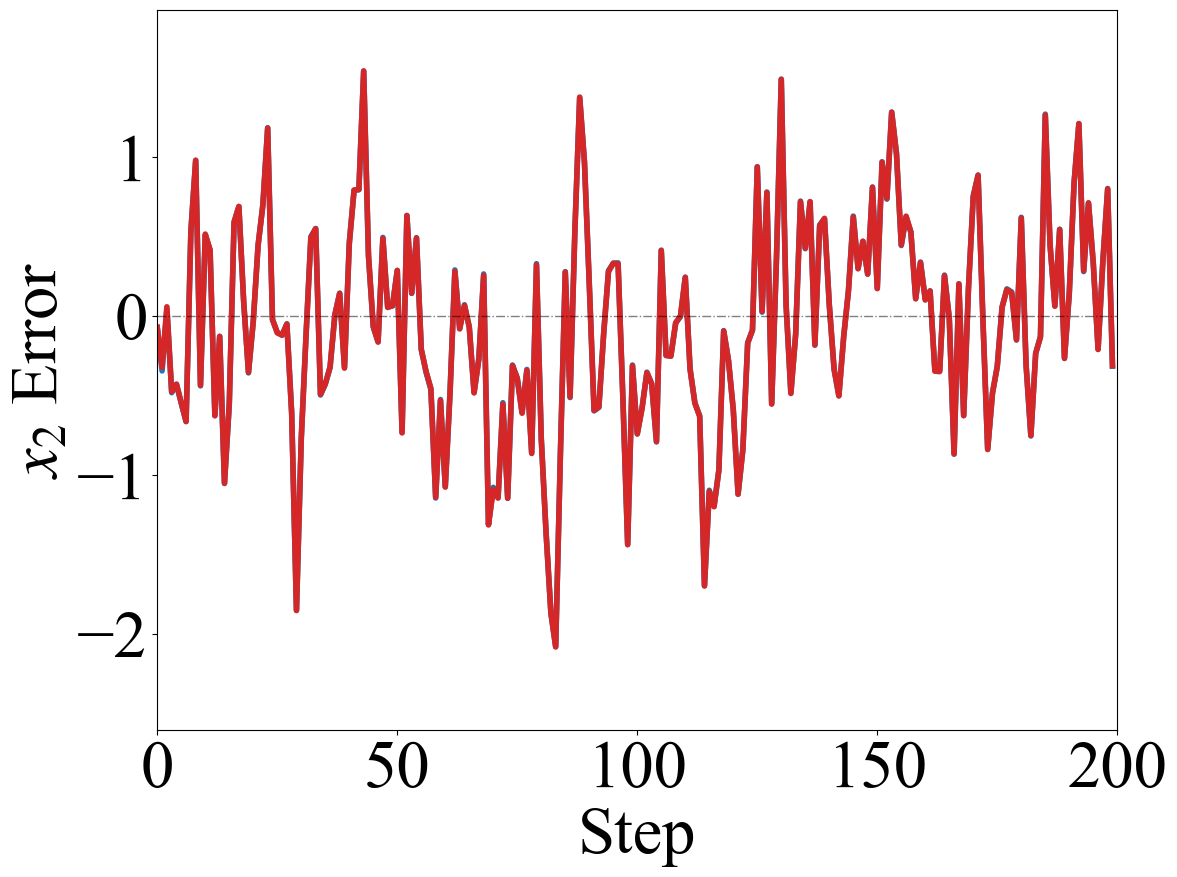}
        \label{osci:2}
    }
    \\
    \subfloat[Case B: Laplace noise]{
        \includegraphics[width=0.225\textwidth]{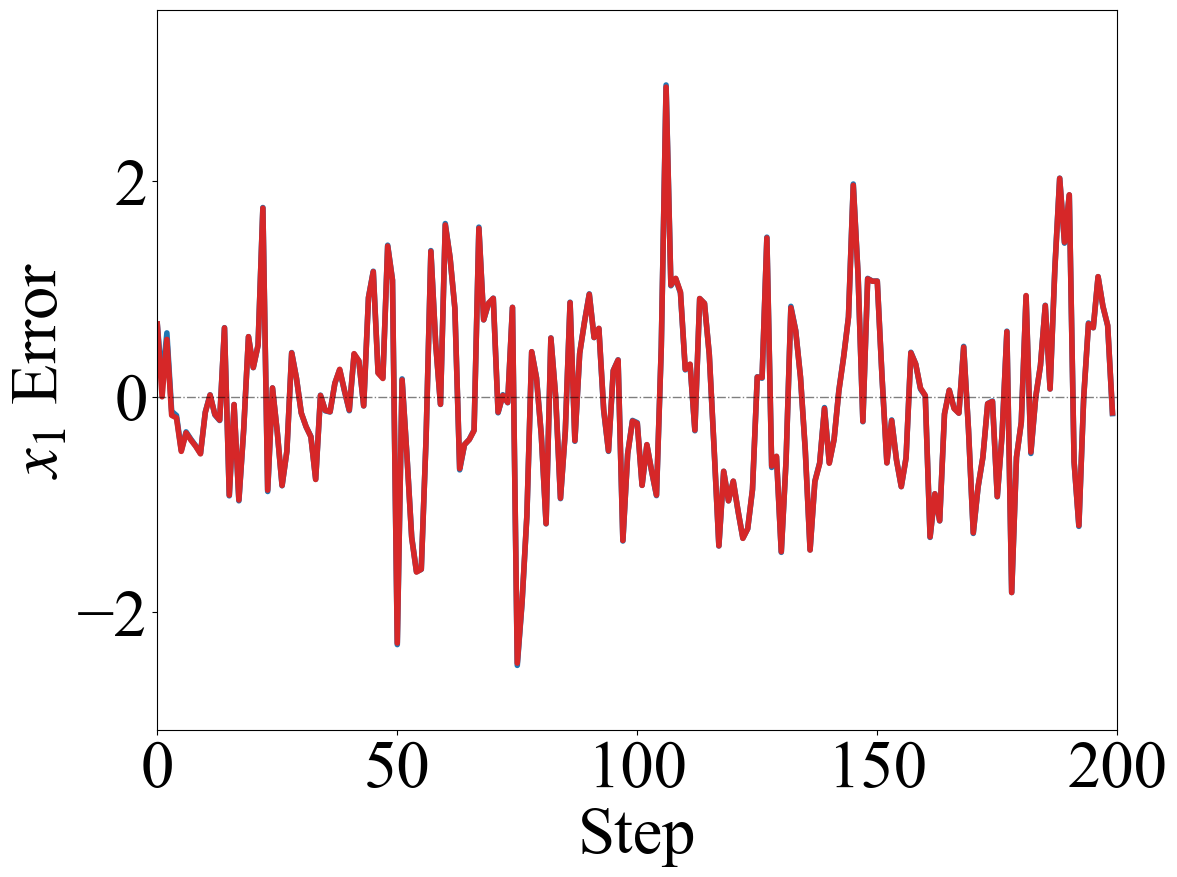}
        \label{osci_la:1}
    \hfill
    \includegraphics[width=0.225\textwidth]{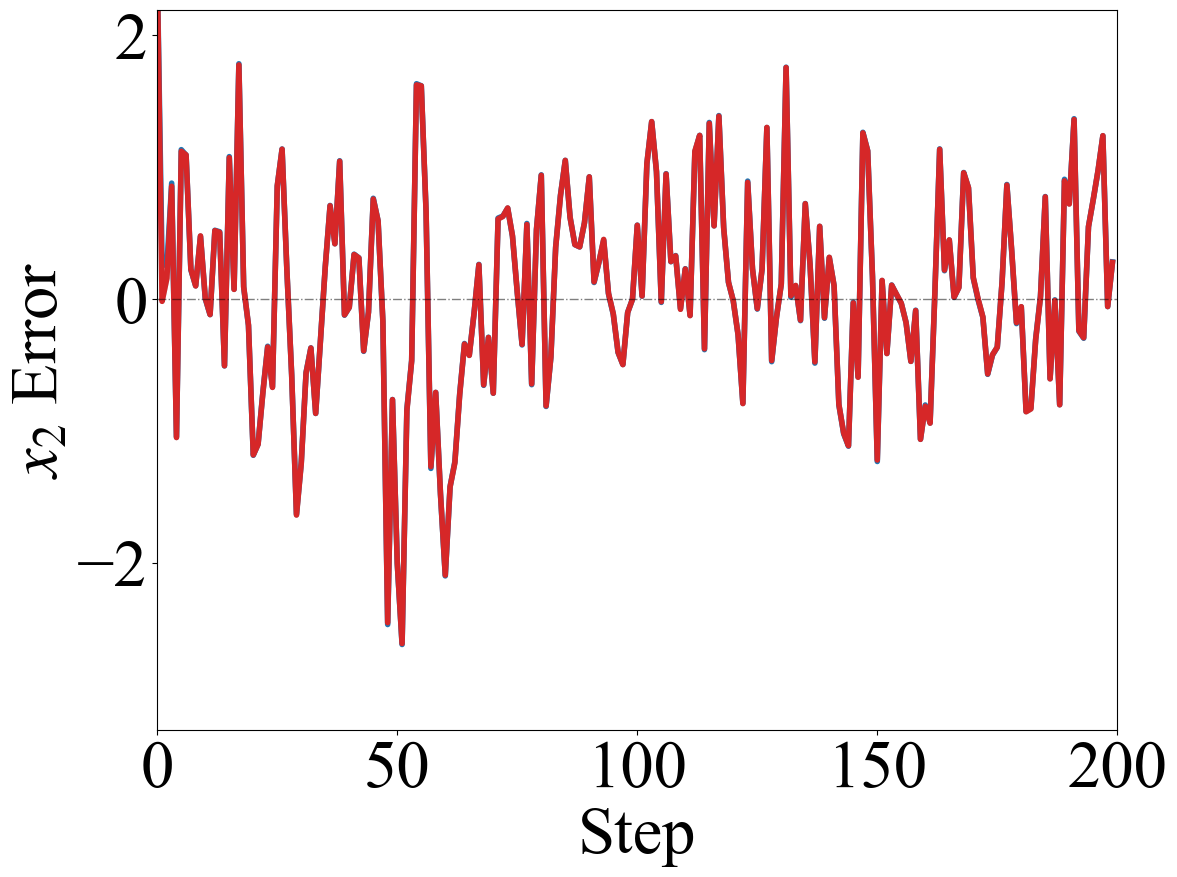}
        \label{osci_la:2}
    }
    \\
    \subfloat[Case C: Beta noise]{
        \includegraphics[width=0.225\textwidth]{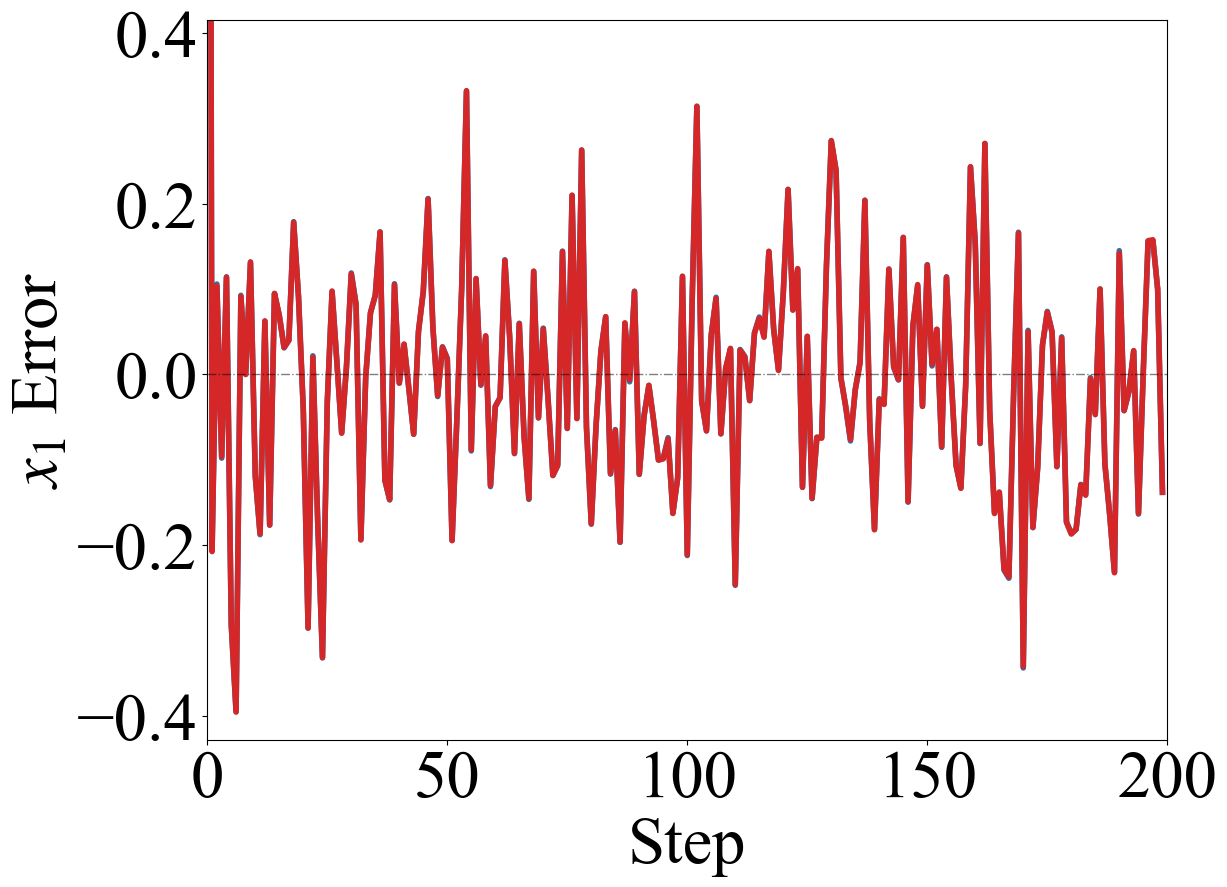}
        \label{osci_beta:1}
    \hfill
    \includegraphics[width=0.225\textwidth]{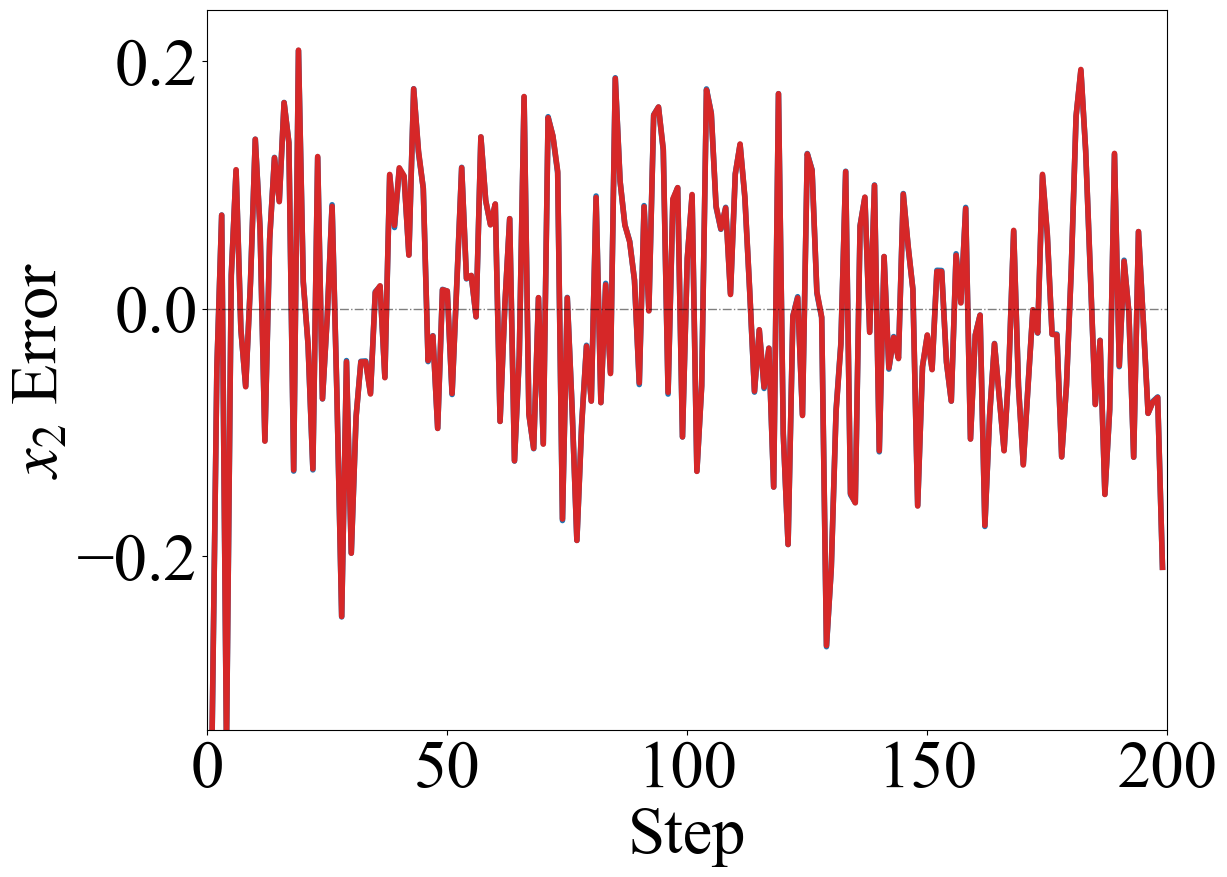}
        \label{osci_beta:2}
    }
    \\
    \subfloat{
        \includegraphics[width=0.45\textwidth]{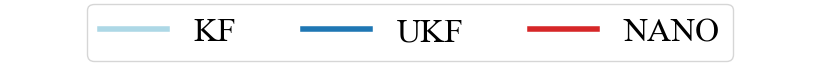}
        \label{osci:legend}
    }
    \caption{Estimation errors in damped linear oscillator.}
    \label{fig.osci_error}
\end{figure} 
In this section, we evaluate our proposed NANO filter across five benchmark systems, including both linear and nonlinear, to demonstrate its applicability over classic filtering algorithms under both Gaussian and non-Gaussian noise conditions. We conduct $N = 100$ Monte Carlo (MC) experiments for each system, so as to avoid the influence of randomness. In each MC experiment, the chosen evaluation metric is the root mean square error (RMSE):
\begin{equation}\nonumber
\text { RMSE }=\sqrt{\frac{1}{M\cdot n} \sum_{t=1}^M\left\|x_t-\hat{x}_t\right\|^2} .
\end{equation}
Here, $x_t, \hat{x}_t$ stand for the real and estimated state, $M$ is the length of running steps and $n$ is the dimension of state. As it is impossible to compare with all existing Gaussian filters, we focus on the most popular ones as baselines for our analysis, including the EKF, UKF, IEKF, and PLF. The algorithms of these filters are all mentioned in the well-regarded textbook \citep{sarkka2023bayesian}.

\subsection{Damped Linear Oscillator}

First, we consider a linear system, named the damped linear  oscillator \citep{course2023state}, which is commonly seen in industrial control field. Its state space model is

\begin{equation}
\nonumber
\begin{aligned}
x_{t+1} &= \exp({\begin{bmatrix}
-0.1 & 2 \\
-2 & -0.1 
\end{bmatrix} \Delta t}) x_t+\xi_t,
\\
y_t &= \begin{bmatrix}
1 & 1 \\
-0.5 & 1
\end{bmatrix} x_t+\zeta_t .
\end{aligned}
\end{equation}
Here, $x_0 \sim \mathcal{N} (\left[2.5, -5\right]^\top, \mathbf{I}_{2\times 2})$ and $\Delta t=0.1$. The process noise $\xi_t$ and measurement noise $\zeta_t$ have three different distributions:

\begin{itemize}
    \item Case A, Gaussian noise: 
    \begin{equation*}
        \xi_t \sim \mathcal{N}(0, 0.5 \cdot \mathbf{I}_{2 \times 2}),  \quad \zeta_t \sim \mathcal{N}(0, \mathbf{I}_{2 \times 2}).
    \end{equation*}
    \item Case B, Laplace noise:
    \begin{equation*}
    \xi_t \sim \text { Laplace }(0, 0.5 \cdot \mathbf{I}_{2 \times 2}),  \quad \zeta_t \sim \text { Laplace }(0, \mathbf{I}_{2 \times 2}).
    \end{equation*}
    \item Case C, Beta noise:
    \begin{equation*}
    \xi_t \sim \text { Beta }(1.5, 2,5) \cdot \mathbf{1}_{2\times 1}, \quad 
    \zeta_t \sim \text { Beta }(2, 5) \cdot \mathbf{1}_{2\times 1}.
    \end{equation*}
\end{itemize}

Fig.~\ref{fig.osci} and Fig.~\ref{fig.osci_error} show the estimation results of damped
linear oscillator system. It can be observed that, under three different noise conditions, the NANO filter achieves the same performance as the KF family, thereby validating the correctness of the NANO filter design.

\subsection{Sequence Forecasting System}
\begin{figure}[!t]
\centering
\subfloat[Case A: Gaussian noise]{
    \includegraphics[width=0.45\textwidth]{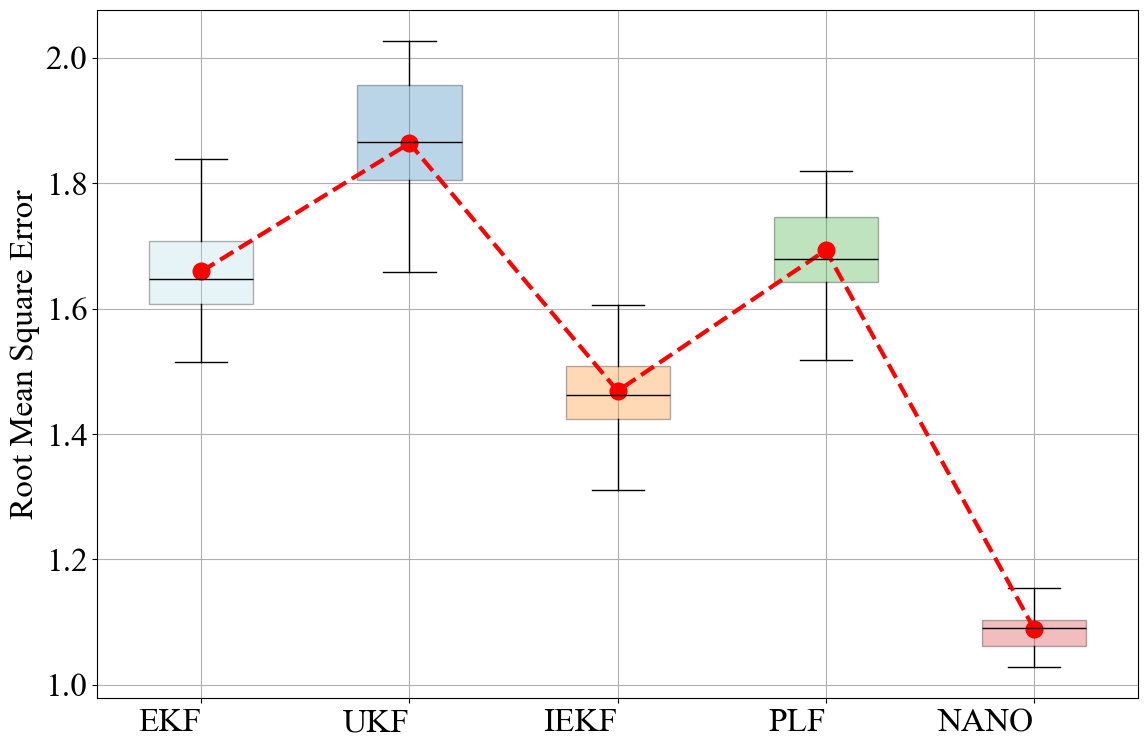}
    \label{fig.sin_cos_gauss}
}
\\
\subfloat[Case B: Laplace noise]{
    \includegraphics[width=0.45\textwidth]{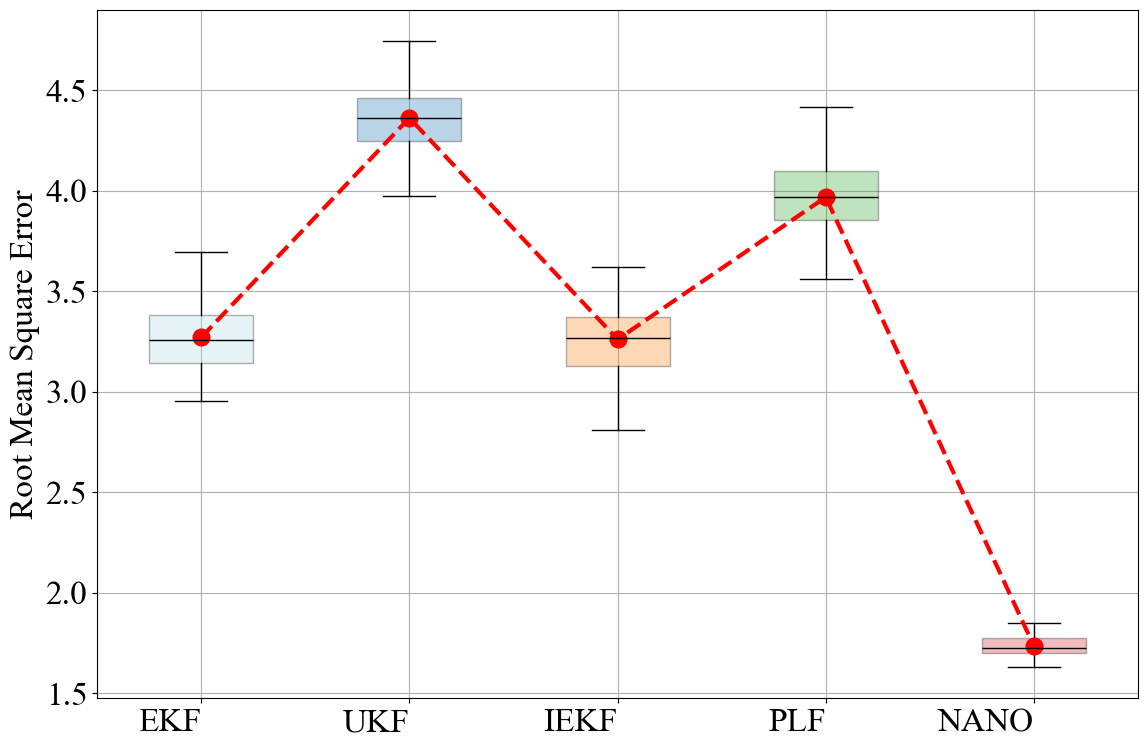}
    \label{fig.sin_cos_la}
}
\\
\subfloat[Case C: Beta noise]{
\includegraphics[width=0.45\textwidth]{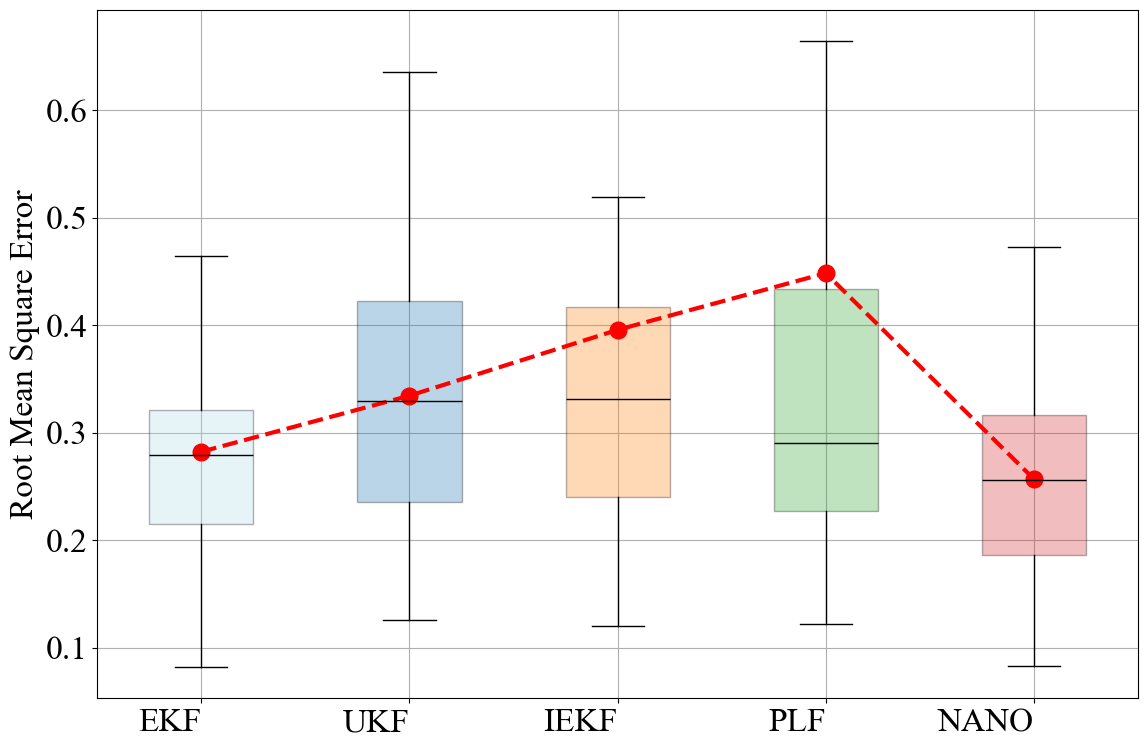}
\label{fig.sin_cos_beta}
}
\caption{Box plot of RMSE over all MC experiments for sequence forecasting
system. Note that red point `` $\textcolor{red}{\bullet}$
 " represents the average RMSE.}
\label{fig.sin_cos}
\end{figure}

\begin{figure}[!t]
    \centering
    \subfloat[Case A: Gaussian noise]{
    \includegraphics[width=0.225\textwidth]{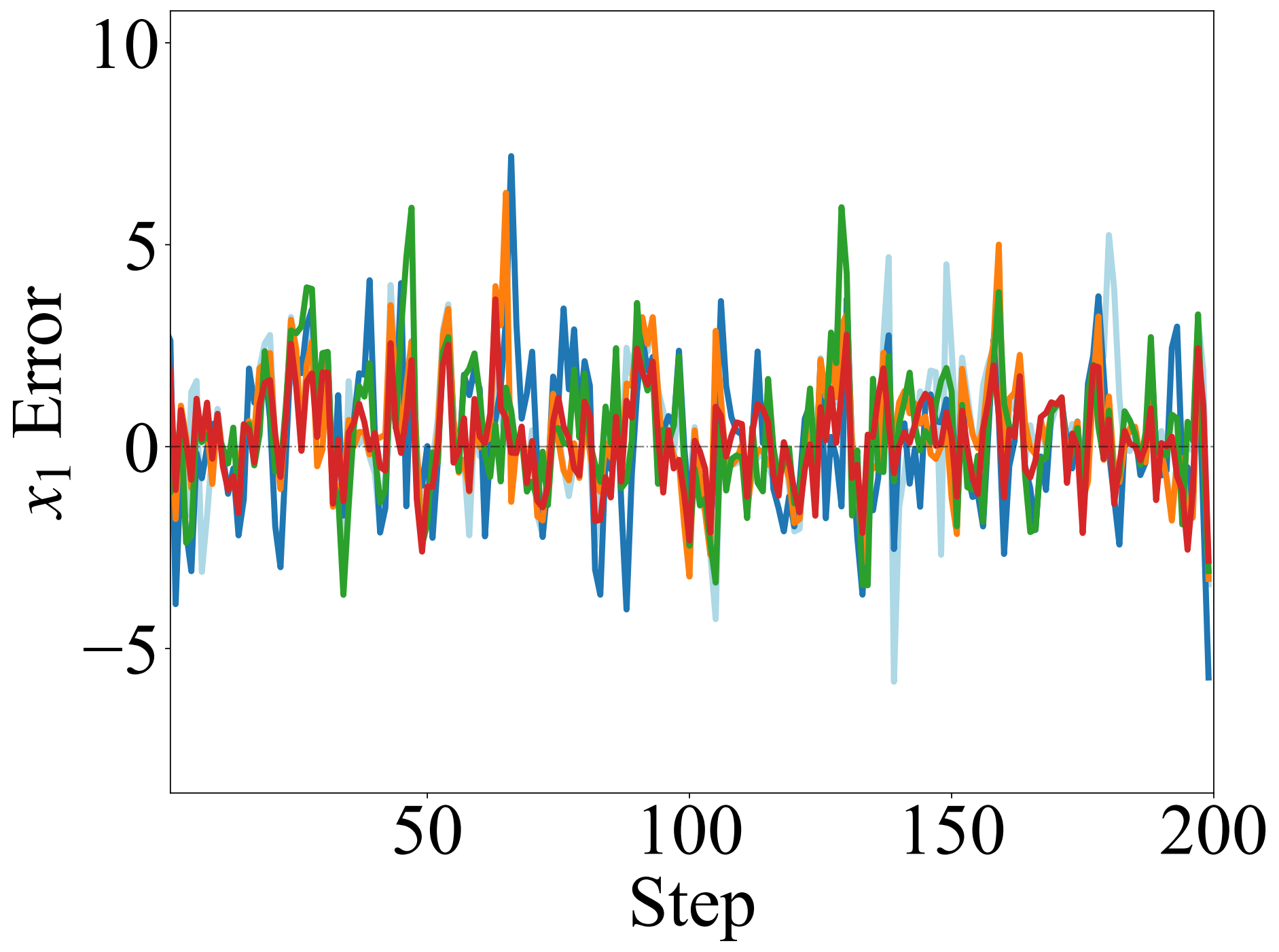}
    \label{sincos:1}
    \hfill
    \includegraphics[width=0.225\textwidth]{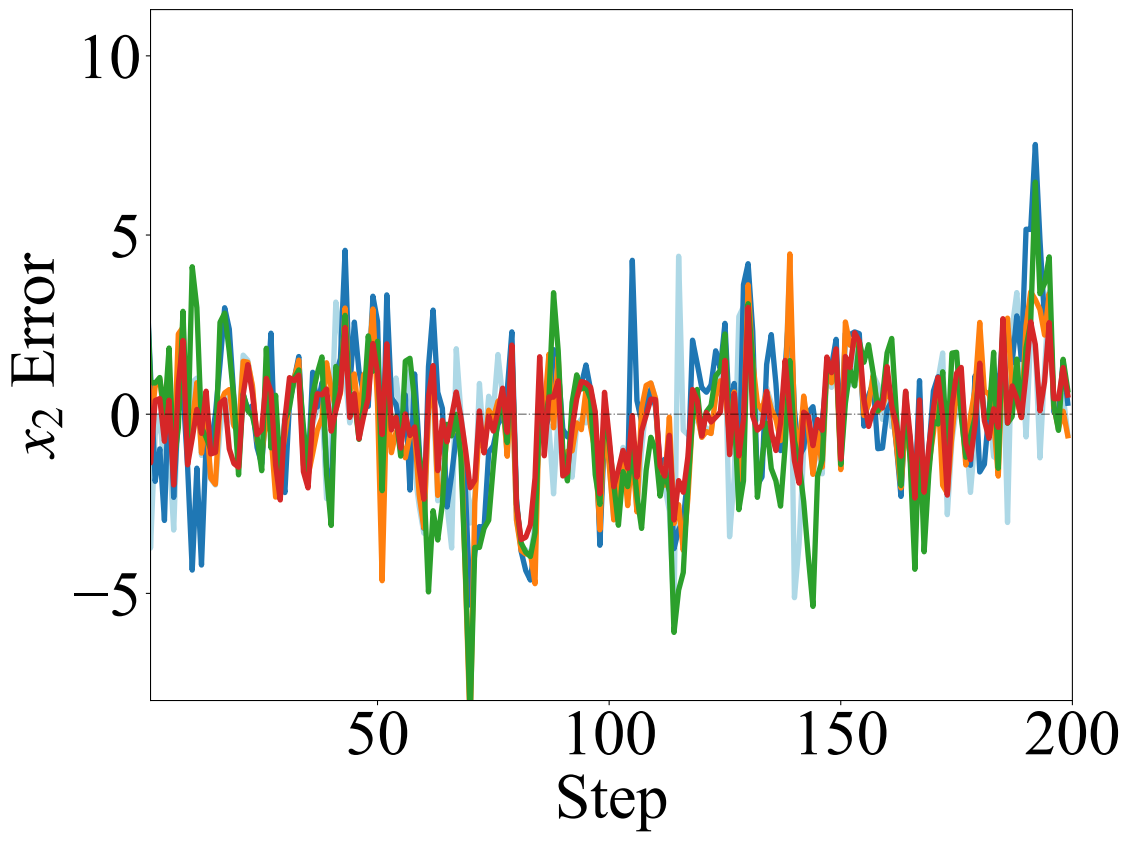}
        \label{sincos:2}
    }
    \\
    \subfloat[Case B: Laplace noise]{
        \includegraphics[width=0.225\textwidth]{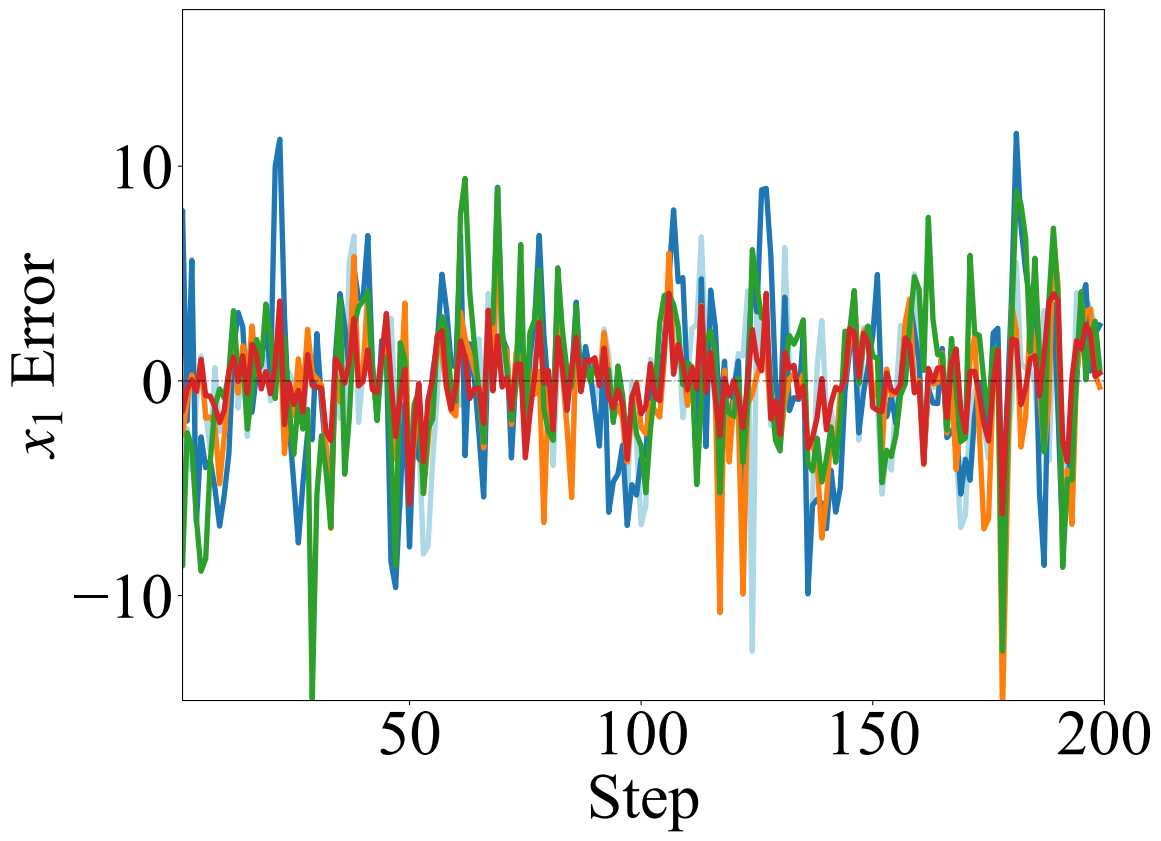}
        \label{sincos_la:1}
    \hfill
    \includegraphics[width=0.225\textwidth]{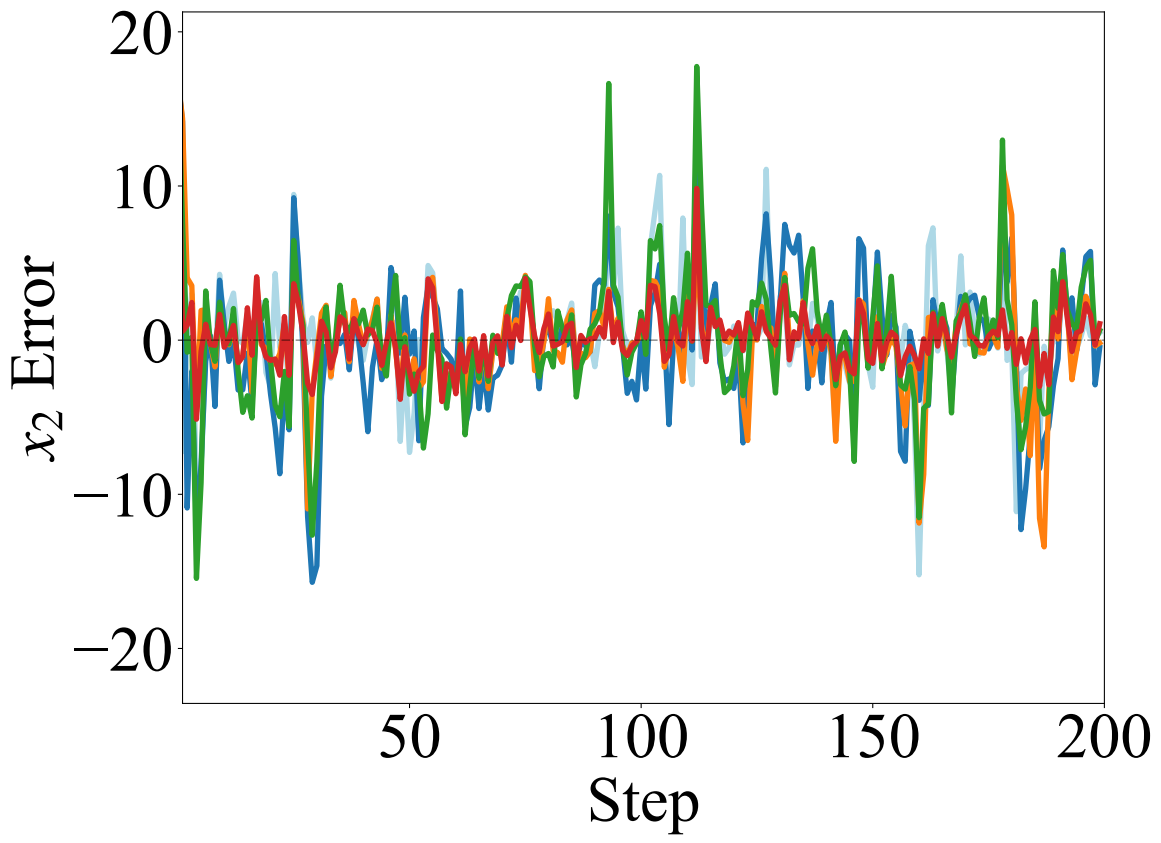}
        \label{sincos_la:2}
    }
    \\
    \subfloat[Case C: Beta noise]{
        \includegraphics[width=0.225\textwidth]{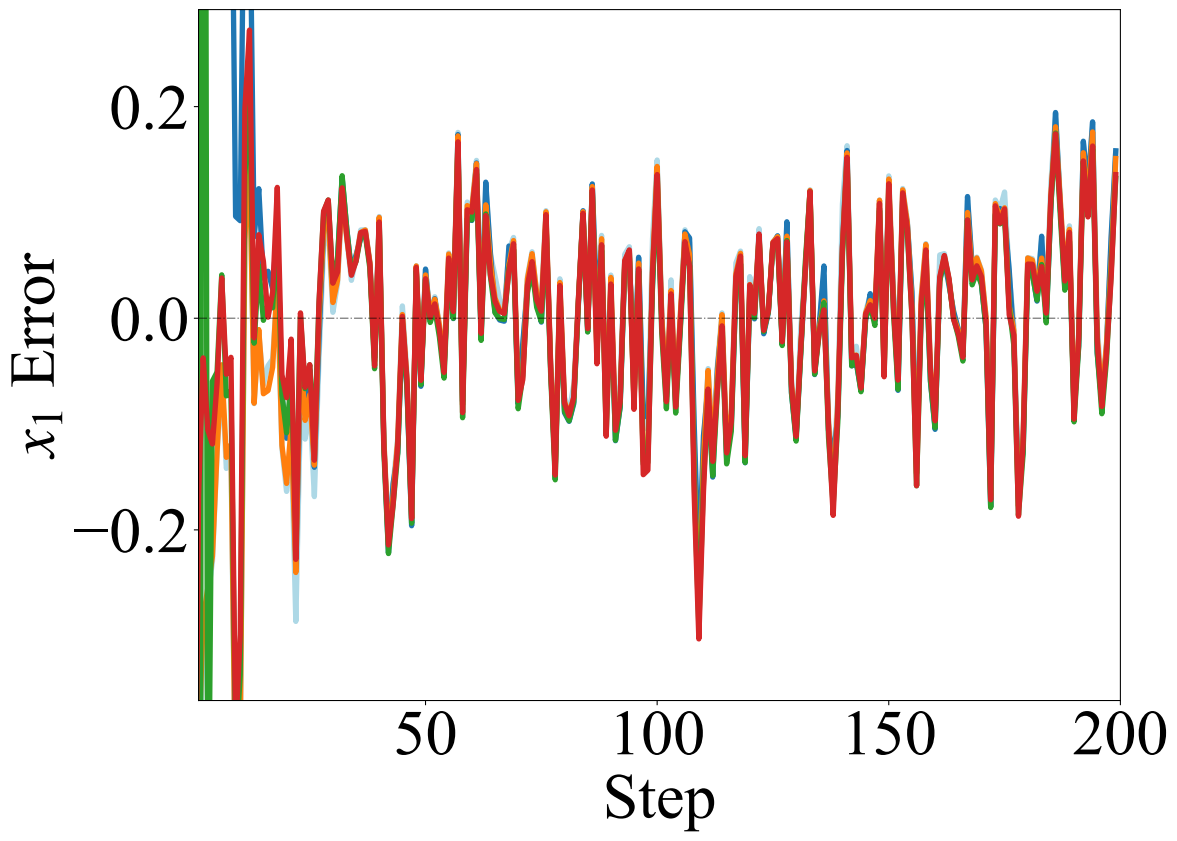}
        \label{sincos_beta:1}
    \hfill
    \includegraphics[width=0.225\textwidth]{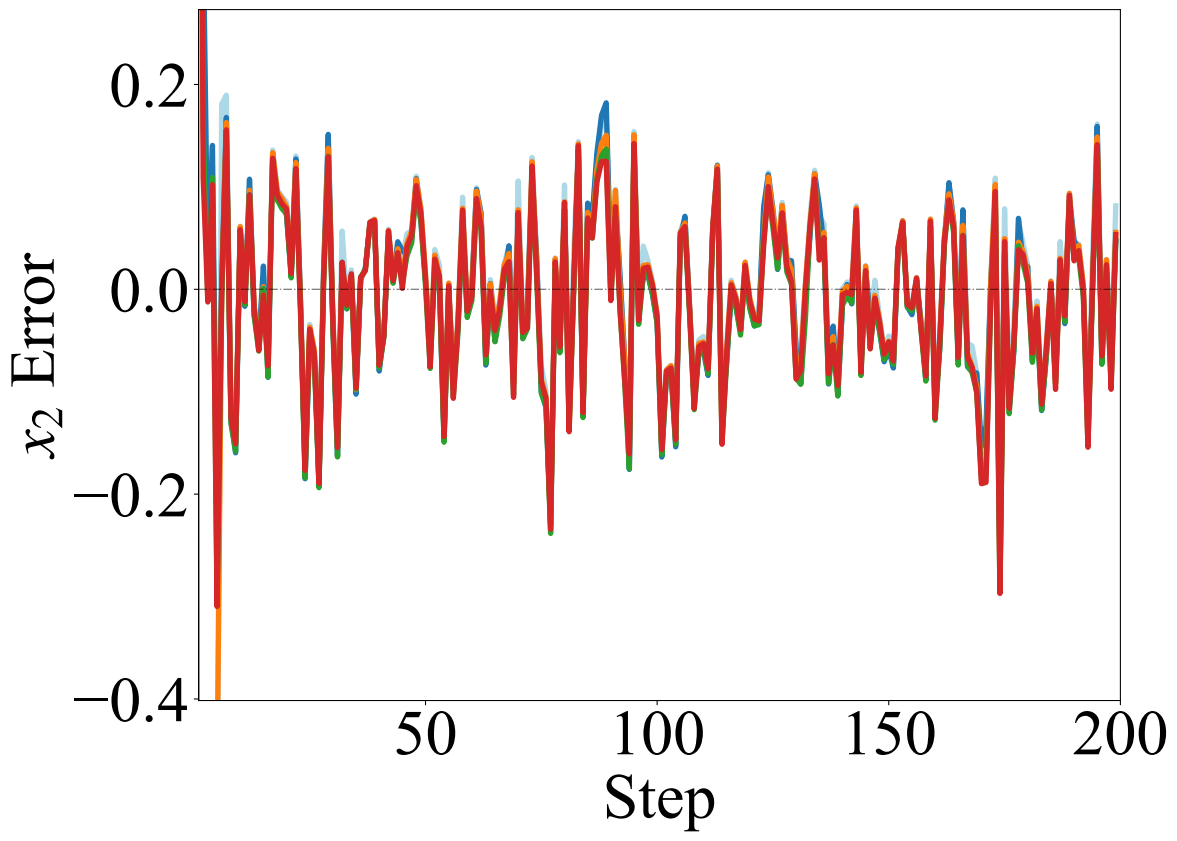}
        \label{sincos_beta:2}
    }
    \\
    \subfloat{
        \includegraphics[width=0.45\textwidth]{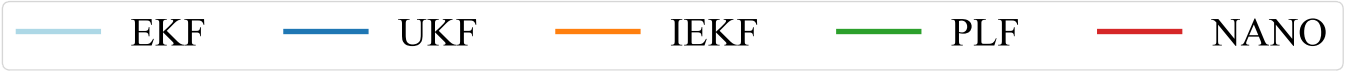}
        \label{sin_cos:legend}
    }
    \caption{ Estimation errors in sequence forecasting system.}
    \label{fig.sin_cos_error}
\end{figure} 
\label{sec.sin_cos}
The following examples are all nonlinear systems, and we first consider a sequence forecasting system \citep{tao2023outlier}. The state space model is given by
\begin{equation}\nonumber
\begin{aligned}
x_{t+1} & =x_t+0.1 \cdot\left[\begin{array}{cc}
-1 & 0 \\
0.1 & -1
\end{array}\right] x_t+0.1 \cdot \cos \left(x_t\right)+\xi_t, \\
y_t & =x_t+\sin \left(x_t\right)+\zeta_t.
\end{aligned}
\end{equation}
Here, $x_0 \sim \mathcal{N}\left(0, \mathbf{I}_{2 \times 2}\right)$. The process noise $\xi_t$ and measurement noise $\zeta_t$ have three different distributions:
\begin{itemize}
    \item Case A, Gaussian noise: 
    \begin{equation*}
        \xi_t \sim \mathcal{N}(0,  4\cdot \mathbf{I}_{2 \times 2}),  \quad \zeta_t \sim \mathcal{N}(0, \mathbf{I}_{2 \times 2}).
    \end{equation*}
    \item Case B, Laplace noise:
    \begin{equation*}
    \xi_t \sim \text { Laplace }(0, 4\cdot\mathbf{I}_{2 \times 2}),  \quad \zeta_t \sim \text { Laplace }(0, \mathbf{I}_{2 \times 2}).
    \end{equation*}
    \item Case C, Beta noise:
    \begin{equation*}
    \xi_t \sim \text { Beta }(1.5, 2) \cdot \mathbf{1}_{2\times 1} , \quad 
    \zeta_t \sim \text { Beta }(3, 7) \cdot \mathbf{1}_{2\times 1}.
    \end{equation*}
\end{itemize}

As shown in Fig.~\ref{fig.sin_cos}, the NANO filter outperforms all other Gaussian filters under three different noise conditions.
Fig.~\ref{fig.sin_cos_error} shows the state estimation error curves for different algorithms. As can be seen, under Gaussian and Laplace noise, the error curve of NANO is closest to zero, significantly outperforming the other filters. Under beta noise, the curves of several filters are quite similar, but NANO still performs slightly better than the other methods.

\subsection{Growth Model}
\begin{figure}[!t]
\centering
\subfloat[Case A: Gaussian noise]{
    \includegraphics[width=0.45\textwidth]{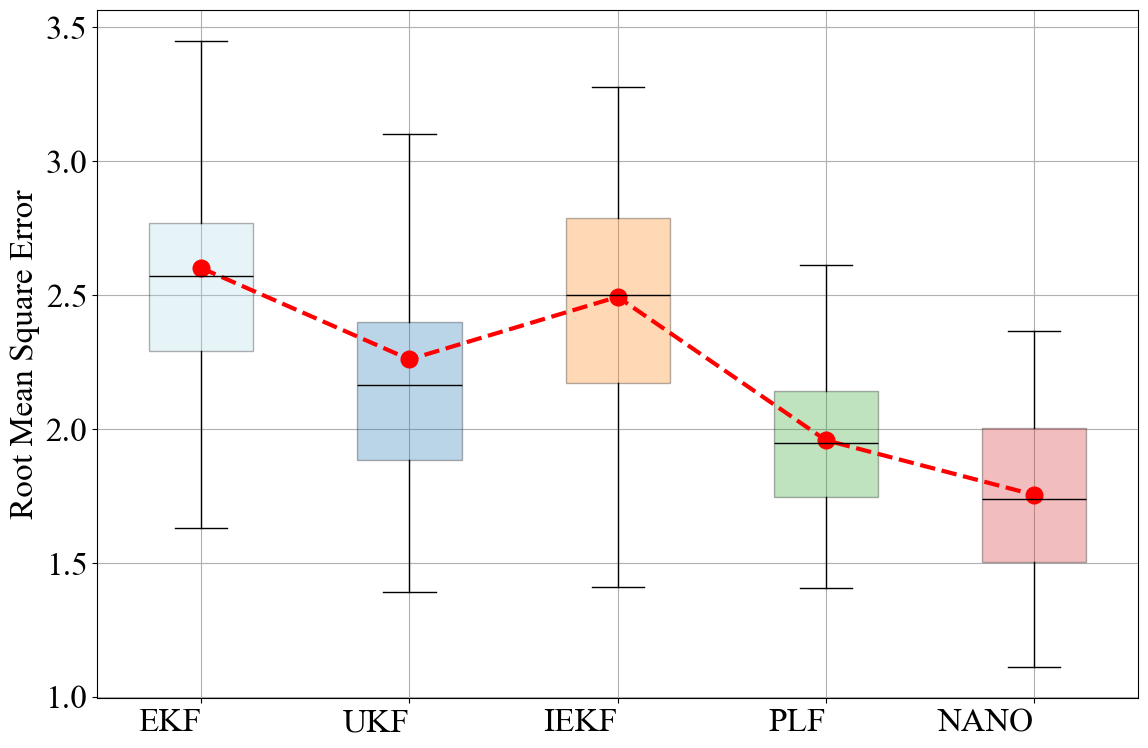}
    \label{fig.toy_gauss}
}
\\
\subfloat[Case B: Laplace noise]{
    \includegraphics[width=0.45\textwidth]{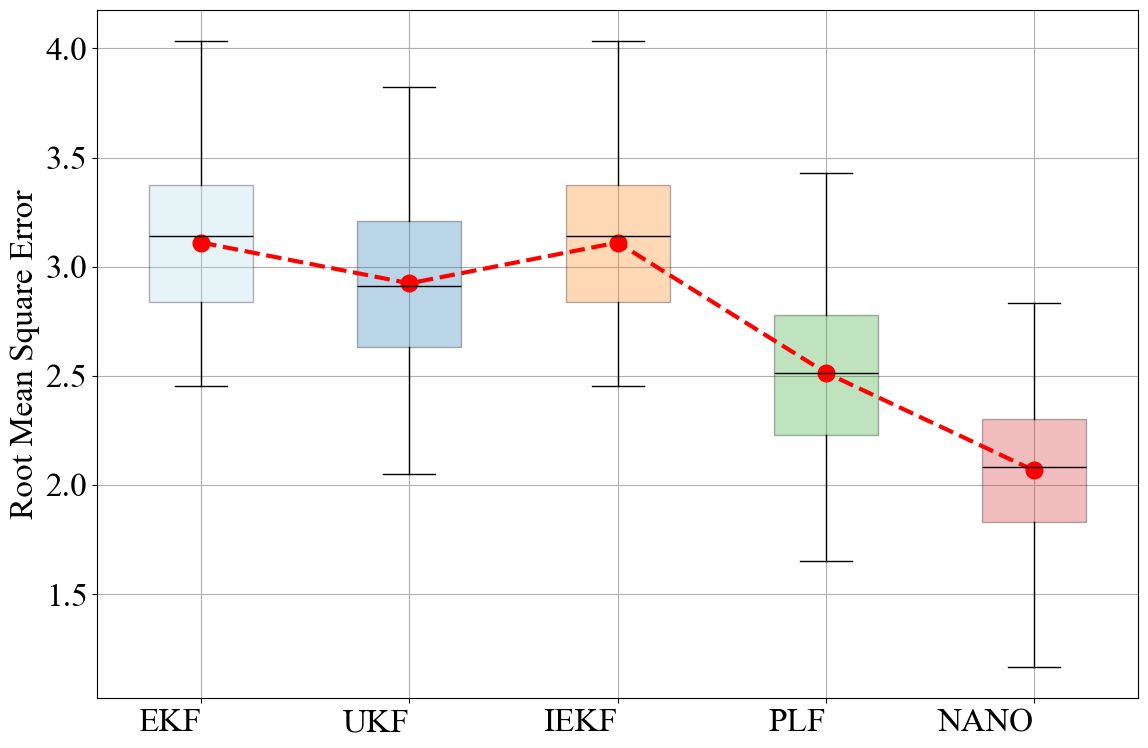}
    \label{fig.toy_la}
}
\\
\subfloat[Case C: Beta noise]{
\includegraphics[width=0.45\textwidth]{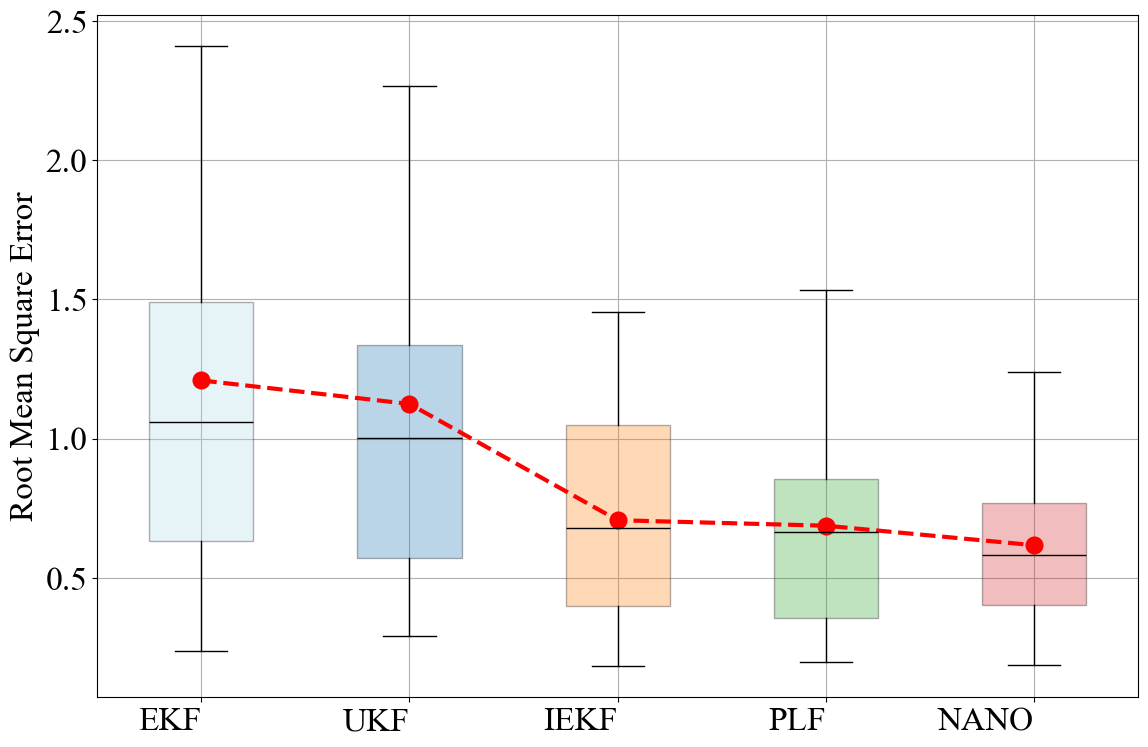}
\label{fig.toy_beta}
}
\caption{Box plot of RMSE over all MC experiments for growth model. Note that red point `` $\textcolor{red}{\bullet}$
 " represents the average RMSE.}
\label{fig.toy}
\end{figure}

\begin{figure}[!t]
    \centering
    \subfloat[Case A: Gaussian noise]{
    \includegraphics[width=0.148\textwidth]{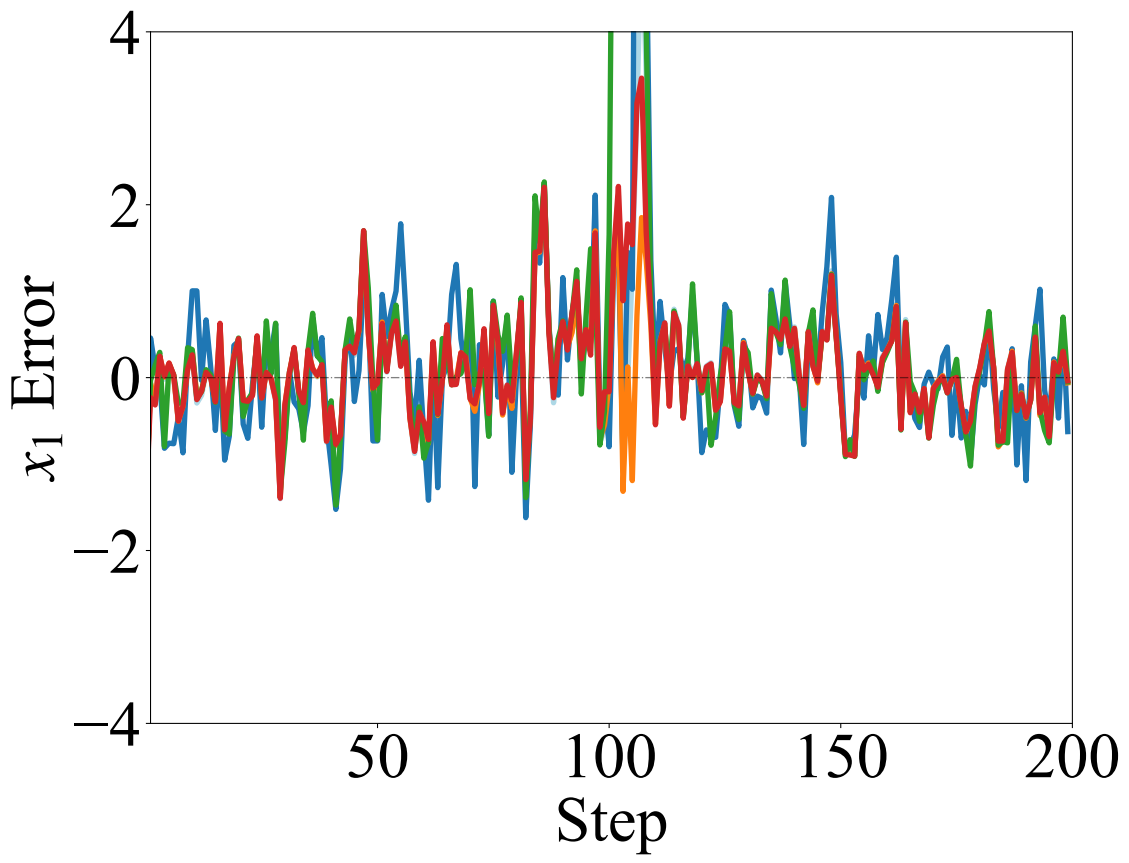}
        \label{toy:1}
    \hfill
    \includegraphics[width=0.148\textwidth]{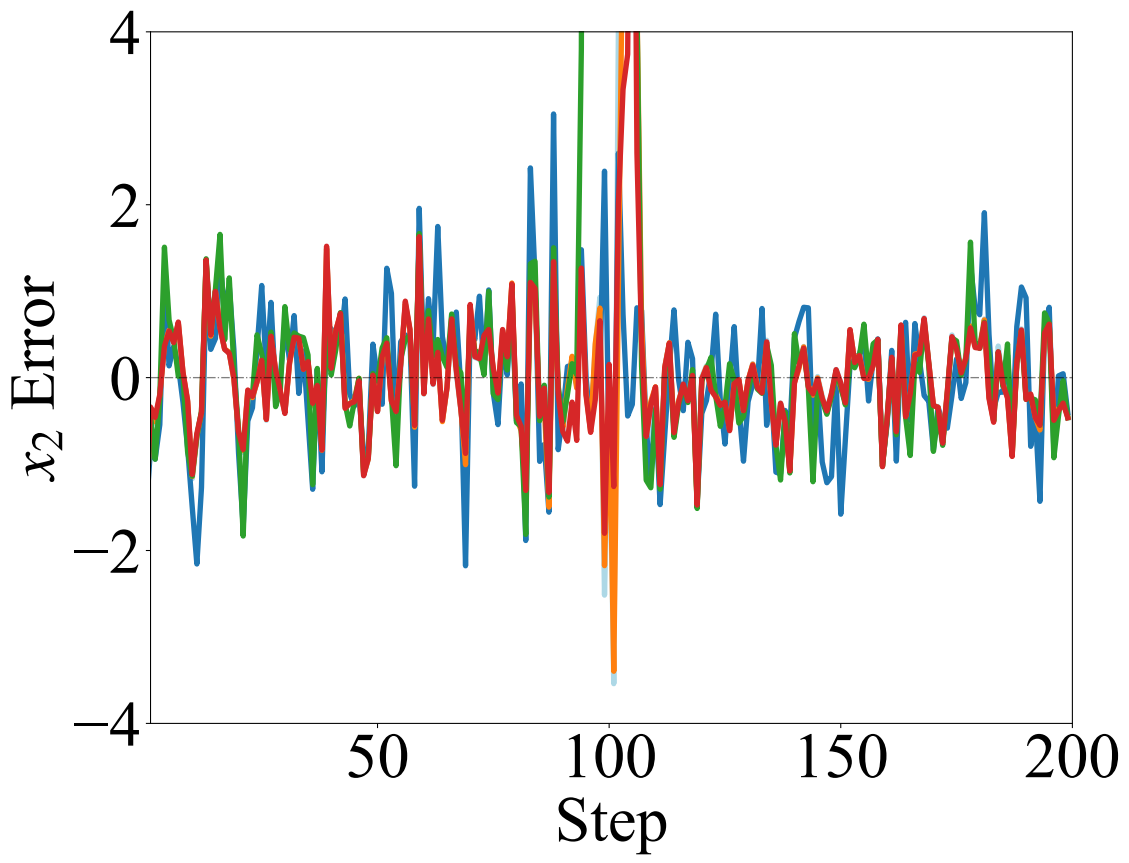}
        \label{toy:2}
    \hfill
    \includegraphics[width=0.148\textwidth]{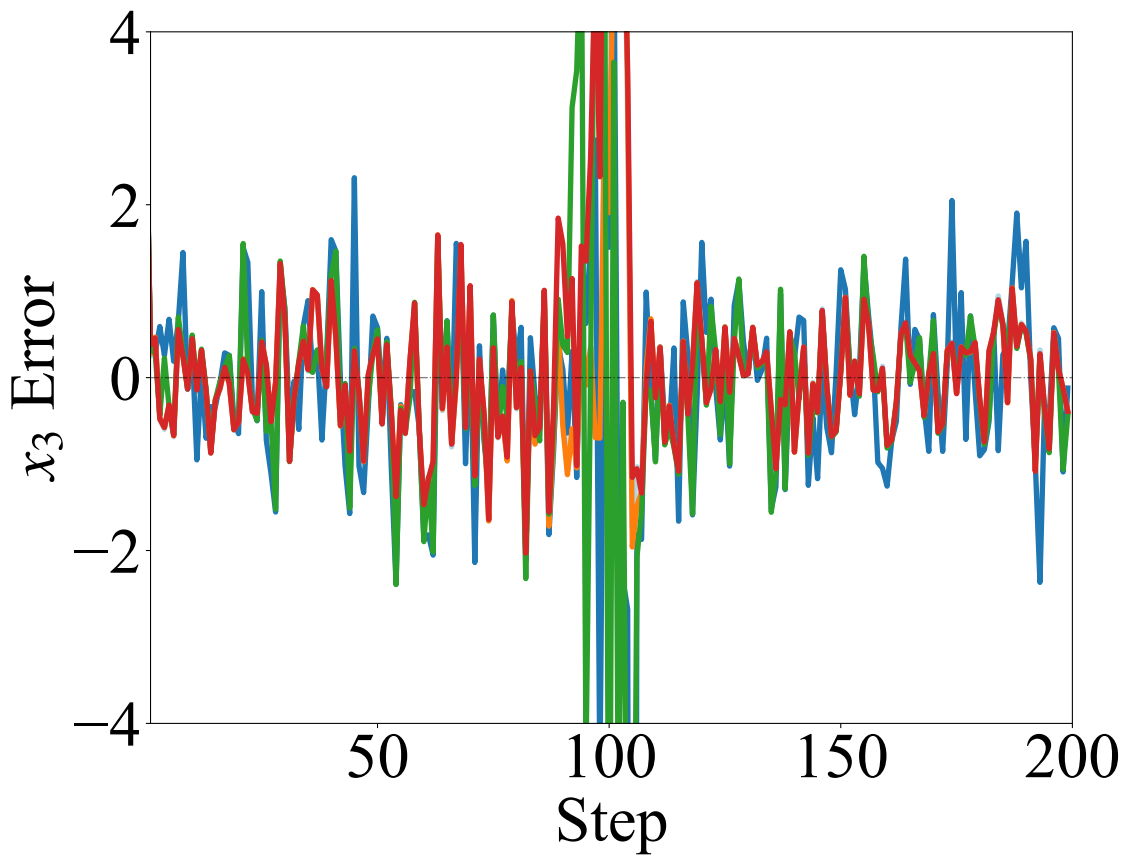}
        \label{toy:3}
    }
    \\
    \subfloat[Case B: Laplace noise]{
    \includegraphics[width=0.148\textwidth]{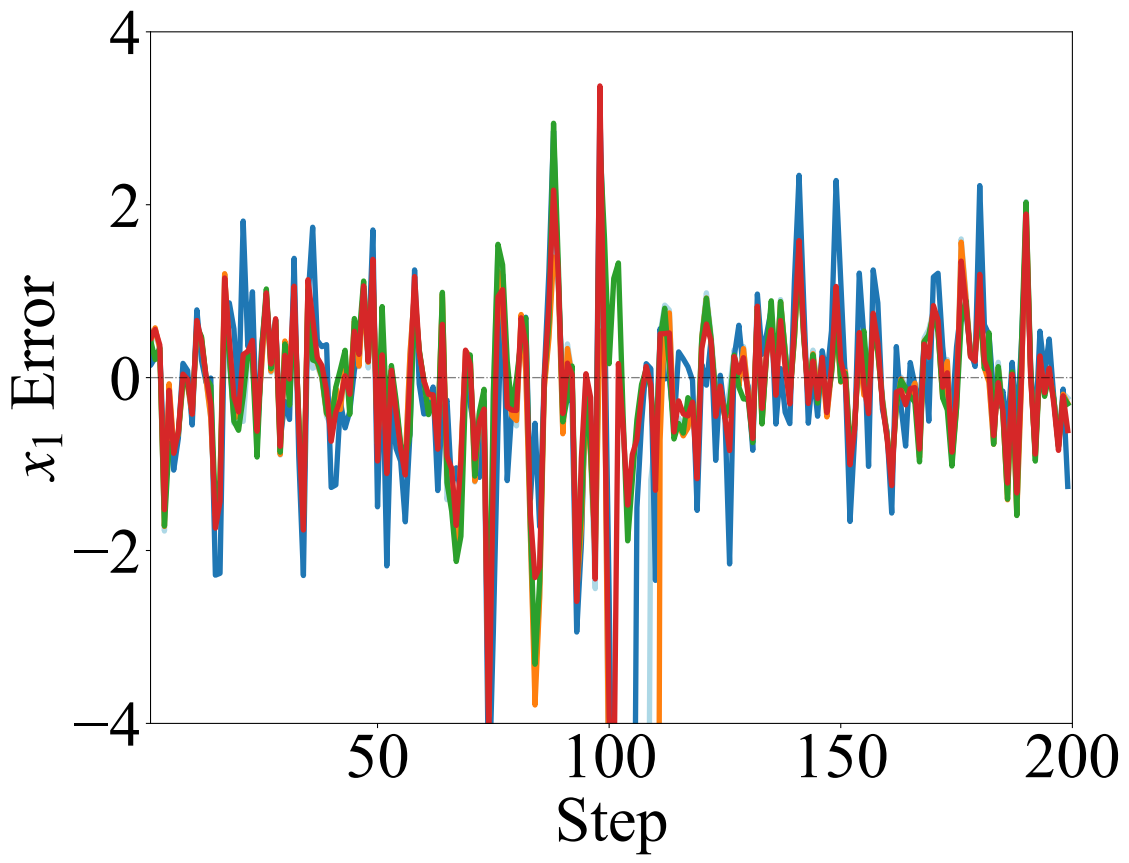}
        \label{toy_la:1}
    \hfill
    \includegraphics[width=0.148\textwidth]{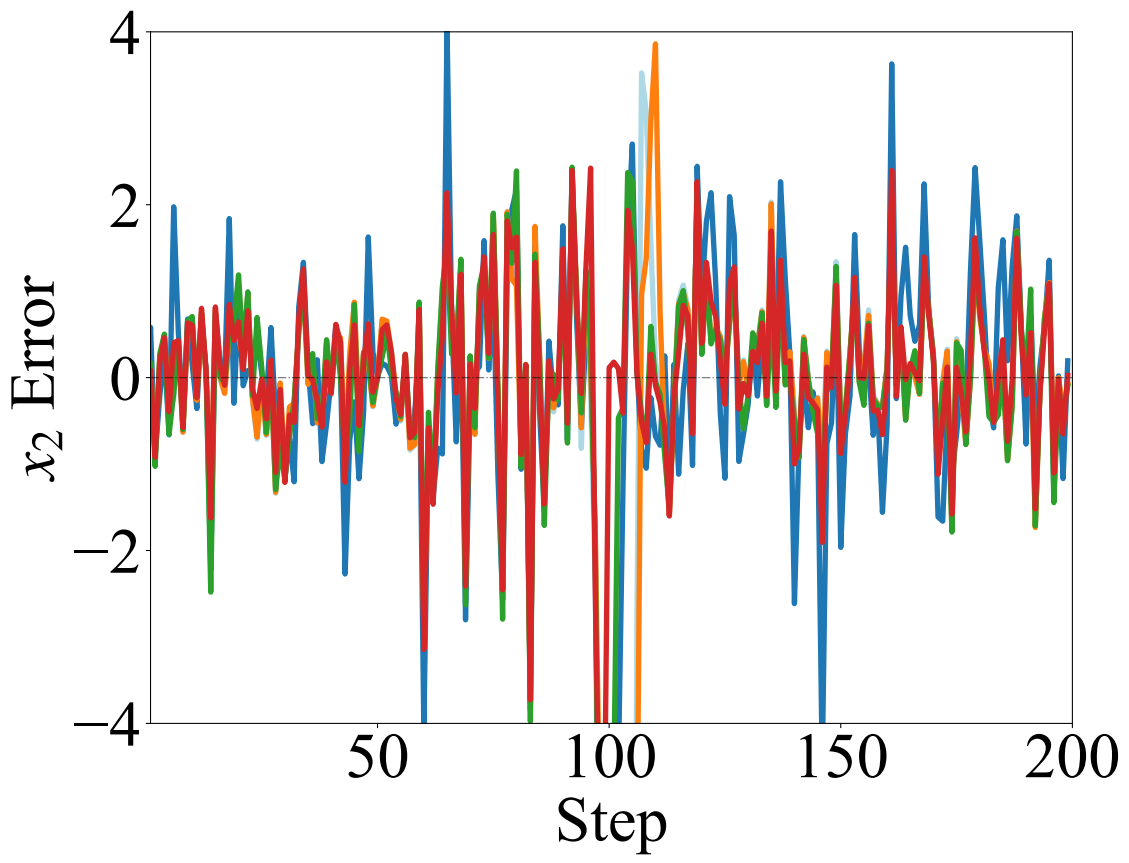}
        \label{toy_la:2}
    \hfill
    \includegraphics[width=0.148\textwidth]{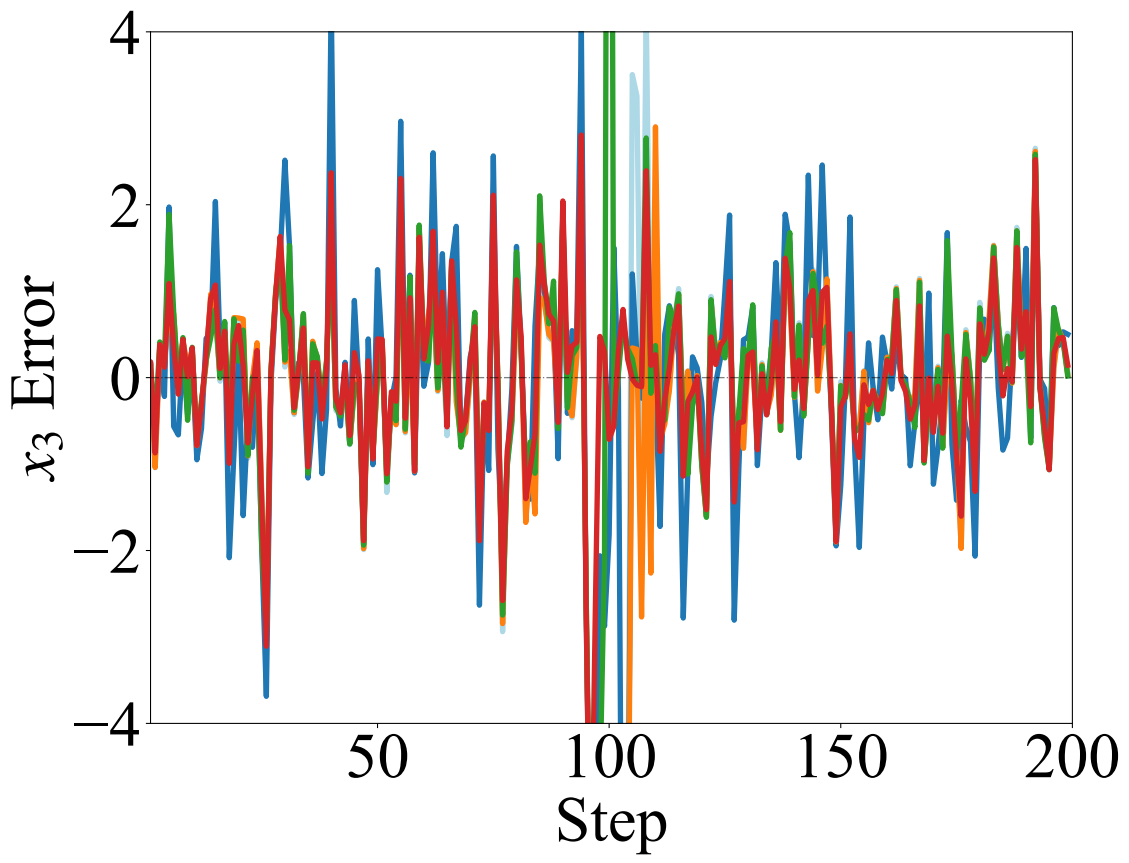}
        \label{toy_la:3}
    }
    \\
    \subfloat[Case C: Beta noise]{
    \includegraphics[width=0.148\textwidth]{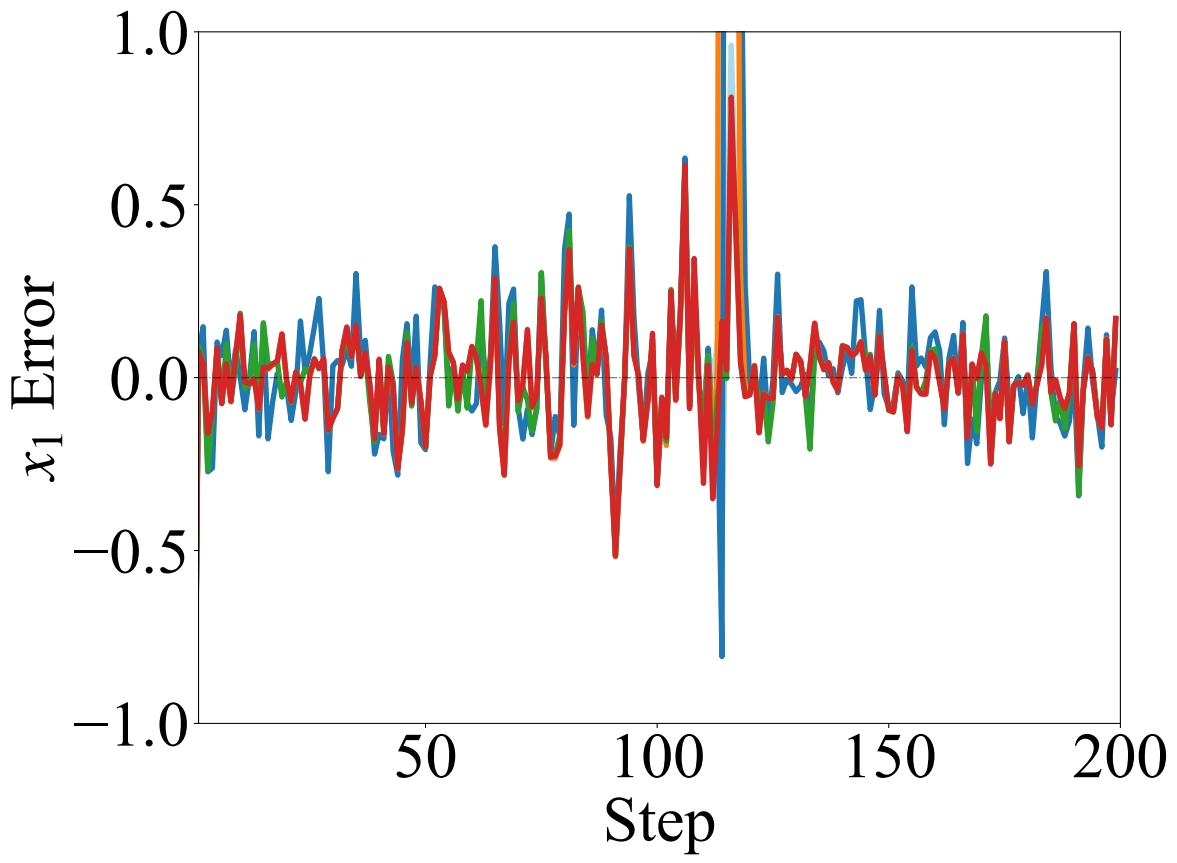}
        \label{toy_beta:1}
    \hfill
    \includegraphics[width=0.148\textwidth]{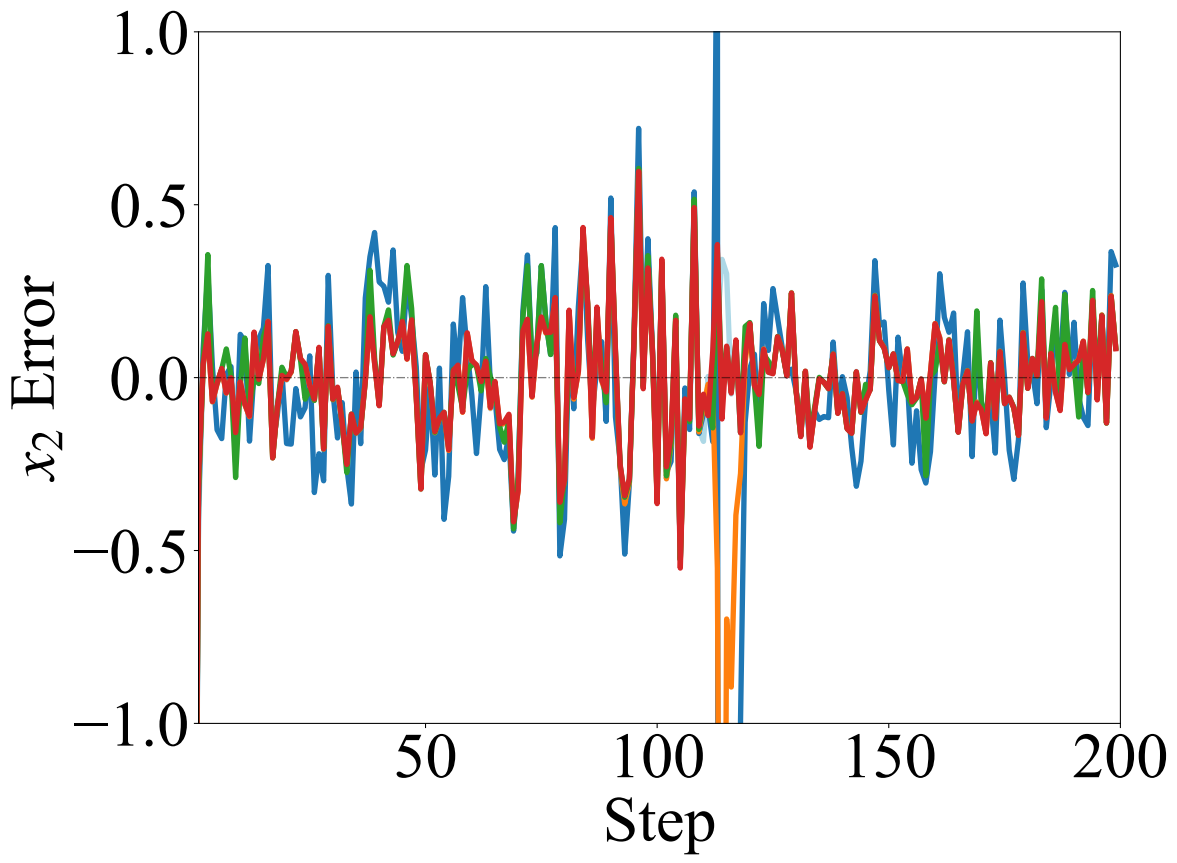}
        \label{toy_beta:2}
    \hfill
    \includegraphics[width=0.148\textwidth]{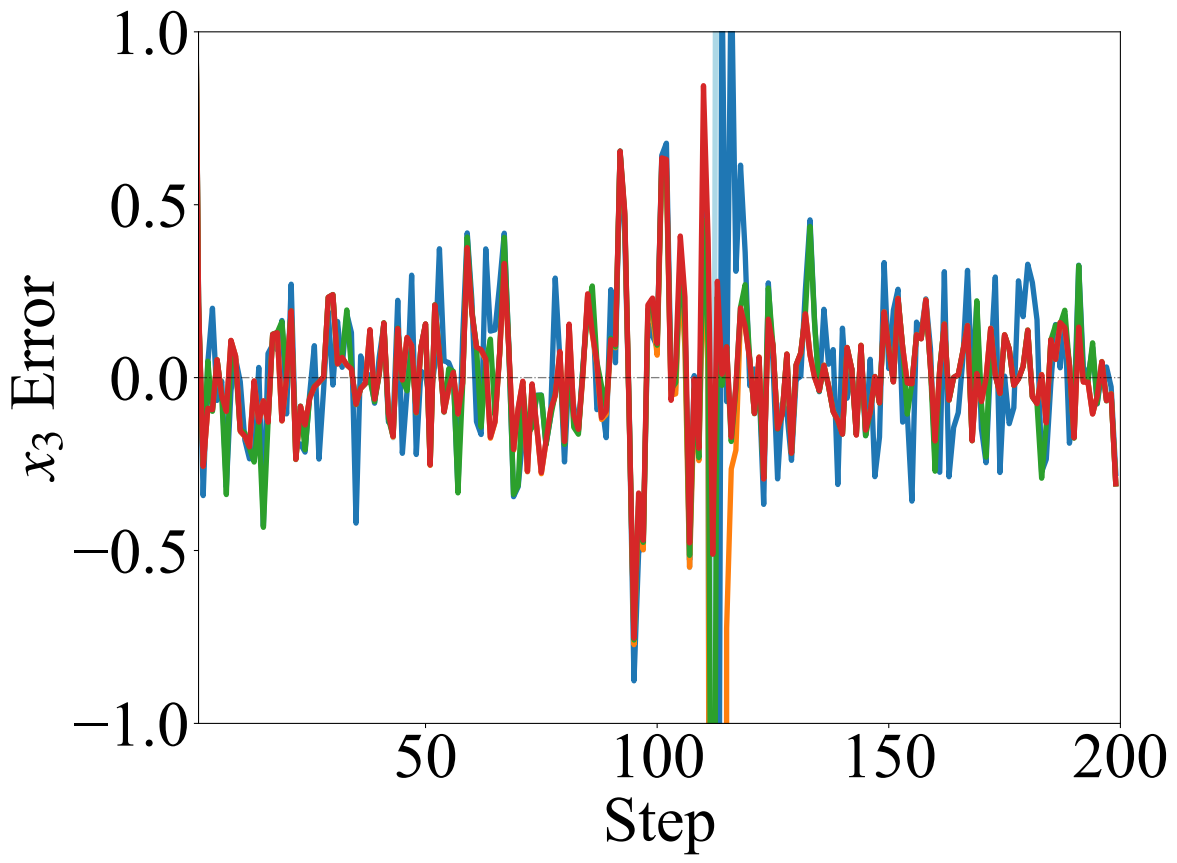}
        \label{toy_beta:3}
    }
    \\
    \subfloat{
        \includegraphics[width=0.45\textwidth]{figures/legend.png}
        \label{toy:legend}
    }
    \caption{Estimation errors in growth model.}
    \label{fig.toy_error}
\end{figure} 
Next, we perform tests on a more strongly nonlinear system called the growth model, which is a benchmark problem that has been used in many literature \citep{gustafsson2010particle, gordon1993novel}. The original dynamic model is modified to equip with coupled nonlinearity, given by
\begin{equation}\nonumber
\begin{aligned}
x_{t+1} & = \begin{bmatrix}
\frac{x_{1,t} + 0.1 x_{2,t}}{2} + \frac{25x_{1,t}}{1 + x_{1,t}^2 + 0.3 x_{2,t}^2} \\
\frac{x_{2,t}+ 0.1 x_{3,t}}{3} + \frac{30x_{2,t}}{1 + x_{2,t}^2 + 0.5 x_{3,t}^2} \\
\frac{x_{3,t}+ 0.1 x_{1,t}}{4} + \frac{35x_{3,t}}{1 + x_{3,t}^2 + 0.7 x_{1,t}^2} 
\end{bmatrix}+ 8\cos(t)\mathbf{1}_{3\times 1}+\xi_t, \\
y_t &= \begin{bmatrix}
\frac{x_{1,t}^2 + x_{2,t}^2}{20} \\
\frac{x_{2,t}^2 + x_{3,t}^2}{20} \\
\frac{x_{1,t}^2 + x_{3,t}^2}{20}
\end{bmatrix} + \zeta_t.
\end{aligned}
\end{equation}
The initial state $x_0 \sim \mathcal{N}(5 \cdot \mathbf{1}_{3\times 1},5\cdot \mathbf{I}_{3\times 3})$. The process noise $\xi_t$ and measurement noise $\zeta_t$ have three different distributions:
\begin{itemize}
    \item Case A, Gaussian noise: 
    \begin{equation*}
        \xi_t \sim \mathcal{N}(0, \mathbf{I}_{3 \times 3}),  \quad \zeta_t \sim \mathcal{N}(0, \mathbf{I}_{3 \times 3}).
    \end{equation*}
    \item Case B, Laplace noise:
    \begin{equation*}
    \xi_t \sim \text { Laplace }(0, \mathbf{I}_{3 \times 3}),  \quad \zeta_t \sim \text { Laplace }(0, \mathbf{I}_{3 \times 3}).
    \end{equation*}
    \item Case C, Beta noise:
    \begin{equation*}
    \xi_t \sim \text { Beta }(2, 2)\cdot \mathbf{1}_{3\times 1} , \quad 
    \zeta_t \sim \text { Beta }(2, 2) \cdot \mathbf{1}_{3\times 1}.
    \end{equation*}
\end{itemize}

The RMSE of state estimation for the growth model using different methods is shown in Fig.~\ref{fig.toy}. Under all noise conditions, the NANO filter consistently achieves a lower average RMSE compared to the other filters, due to its superior effectiveness in handling nonlinear dynamics. 
As shown in Fig.~\ref{fig.toy_error}, the error curve of the NANO filter is the smoothest, with fewer and smaller error spikes.

\subsection{Robot Localization}
\begin{figure}[!t]
\centering
\subfloat[Case A: Gaussian noise]{
    \includegraphics[width=0.45\textwidth]{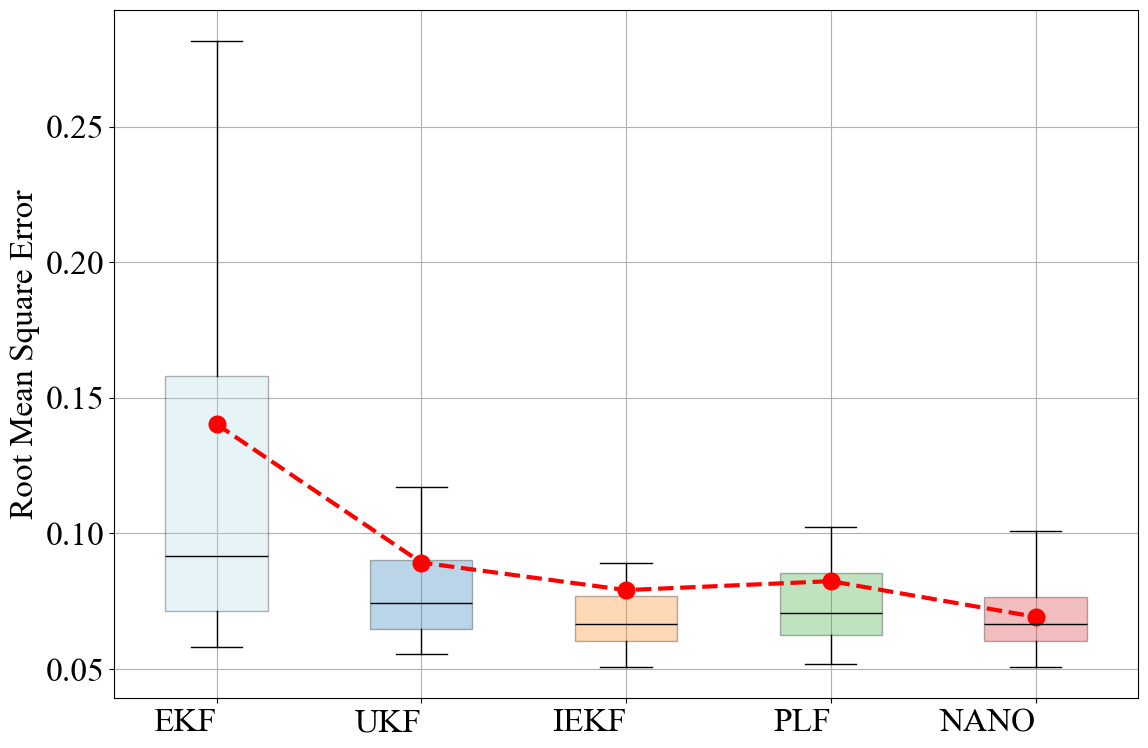}
    \label{fig.localization_gauss}
}
\\
\subfloat[Case B: Laplace noise]{
    \includegraphics[width=0.45\textwidth]{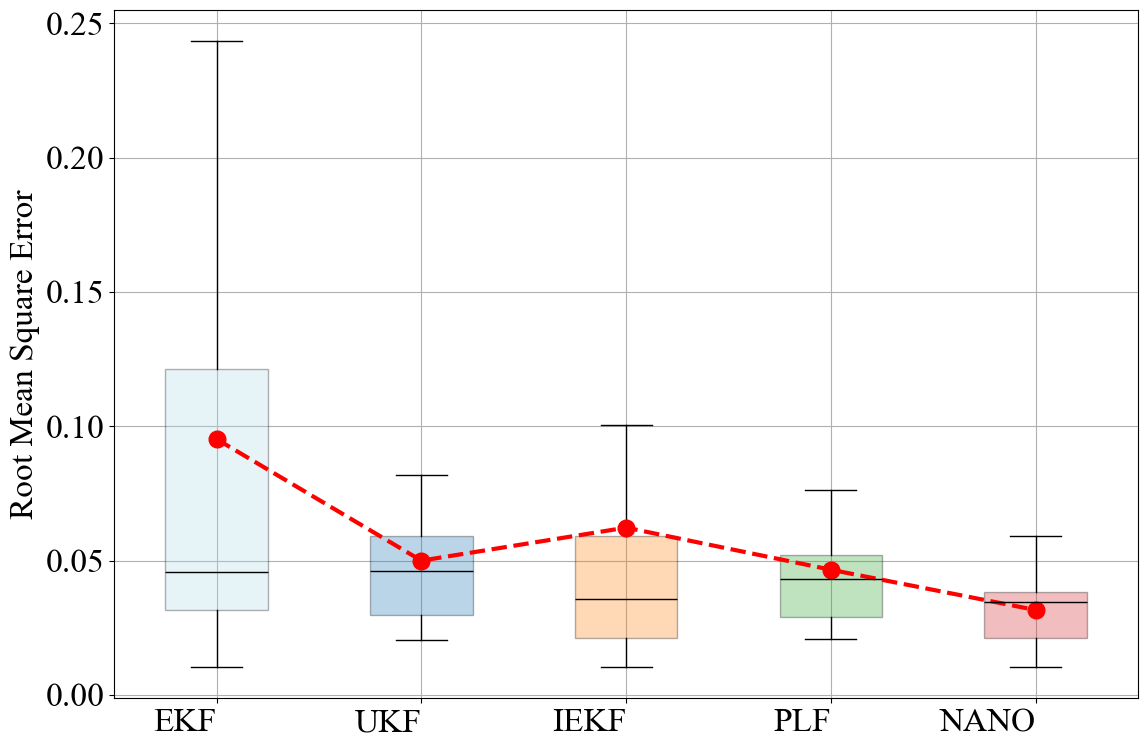}
    \label{fig.localization_la}
}
\\
\subfloat[Case C: Beta noise]{
\includegraphics[width=0.45\textwidth]{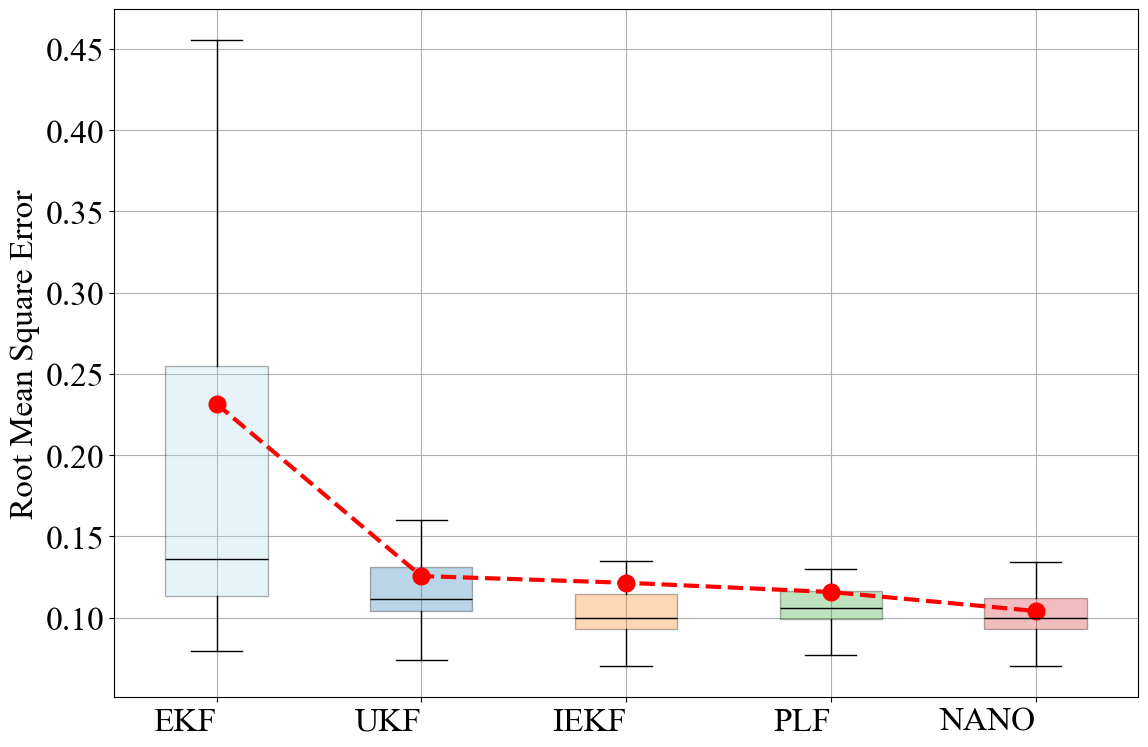}
\label{fig.localization_beta}
}
\caption{Box plot of RMSE over all MC experiments for robot localization. Note that red point `` $\textcolor{red}{\bullet}$
 " represents the average RMSE.}
\label{fig.localization}
\end{figure}

\begin{figure}[!t]
    \centering
    \subfloat[Case A: Gaussian noise]{
    \includegraphics[width=0.148\textwidth]{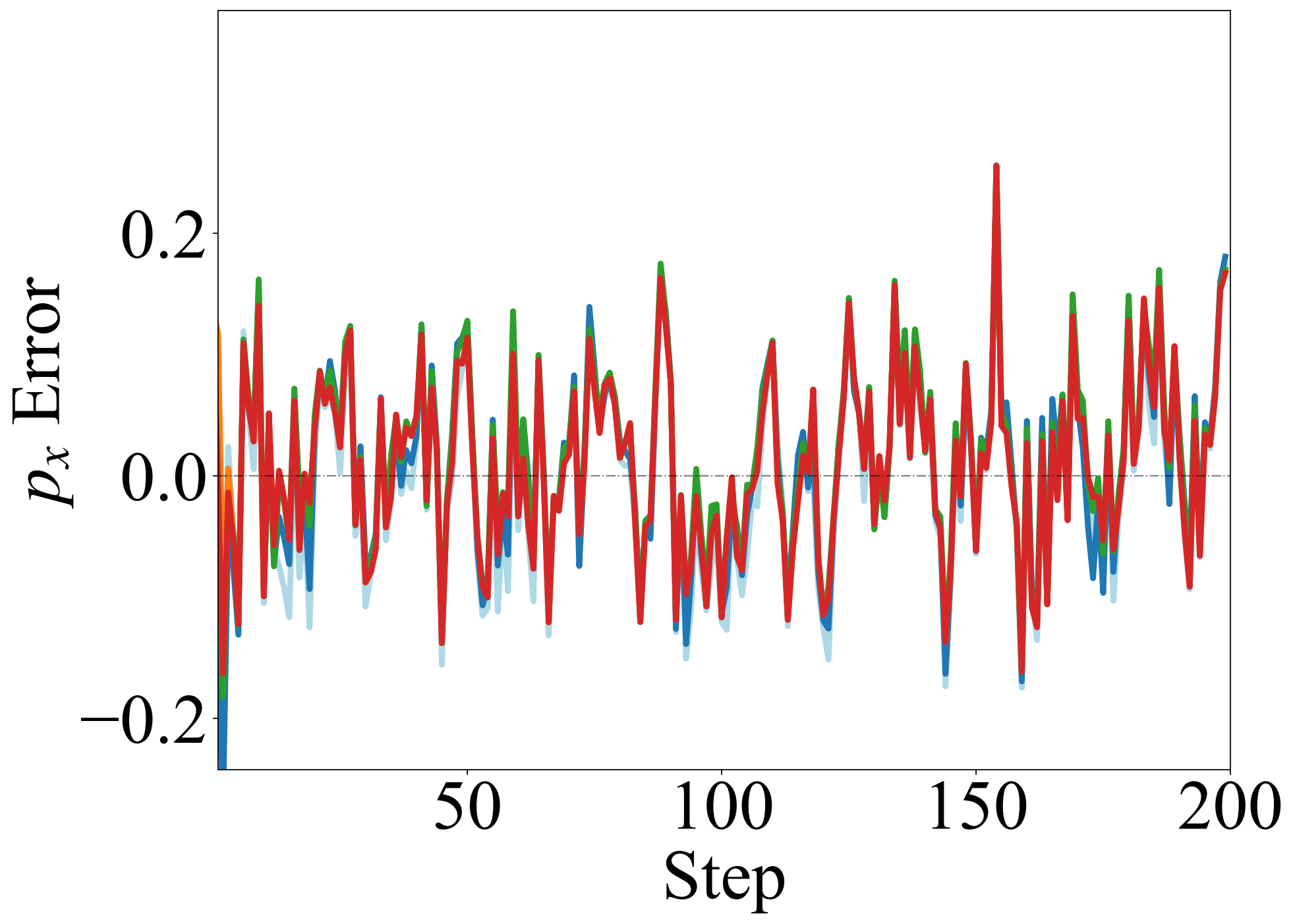}
        \label{localization:1}
    \hfill
    \includegraphics[width=0.148\textwidth]{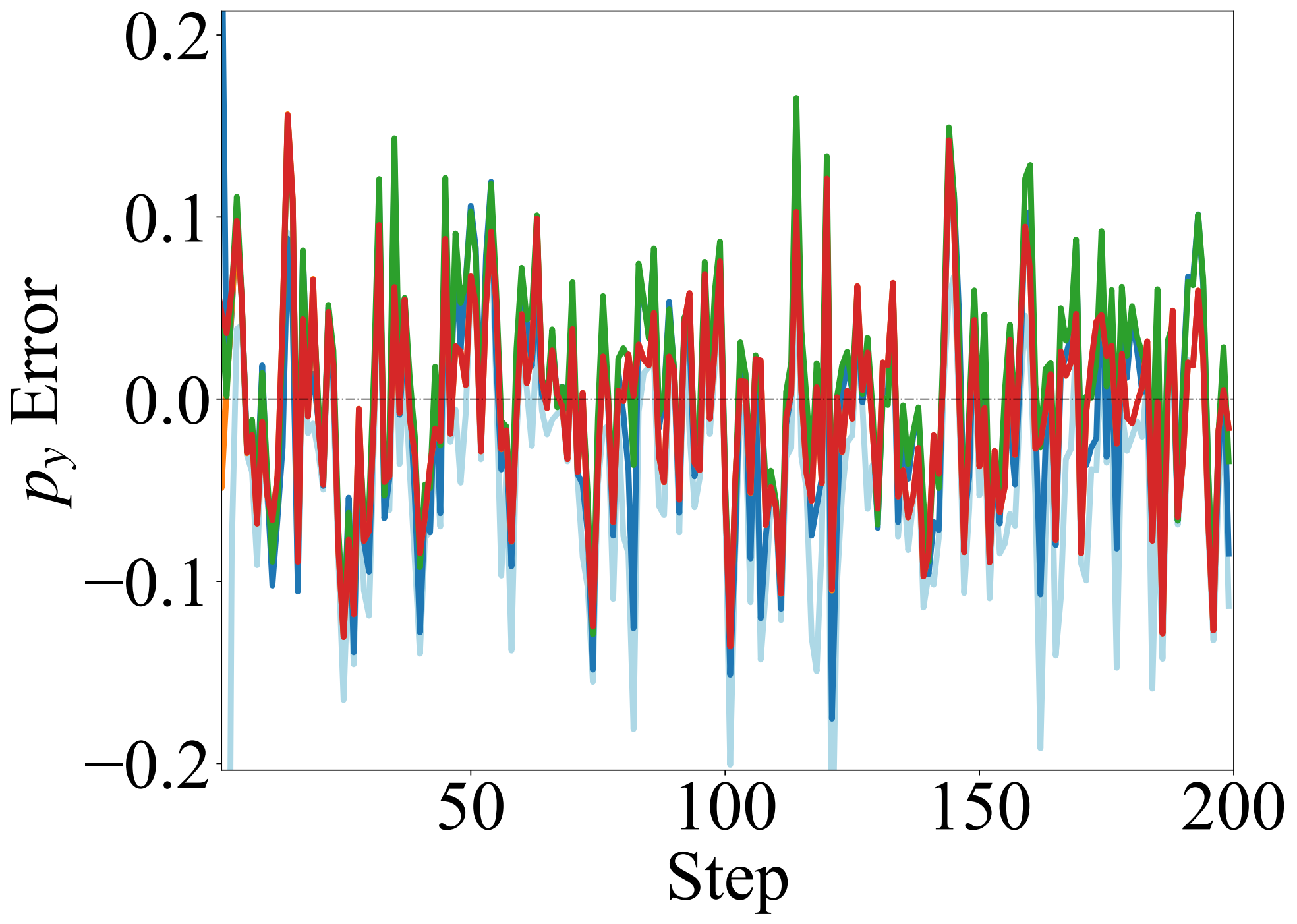}
        \label{localization:2}
    \hfill
    \includegraphics[width=0.148\textwidth]{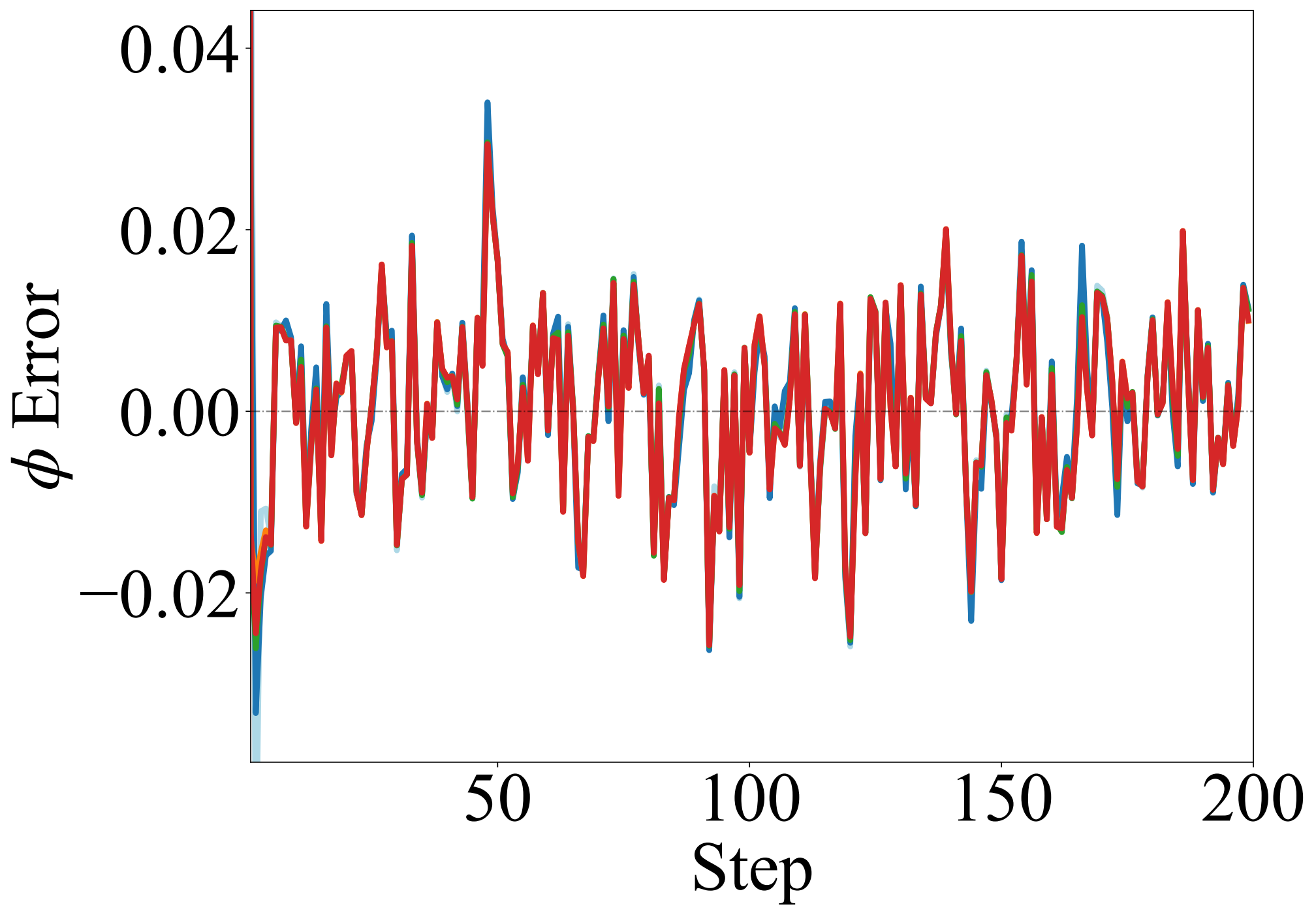}
        \label{localization:3}
    }
    \\
    \subfloat[Case B: Laplace noise]{
    \includegraphics[width=0.148\textwidth]{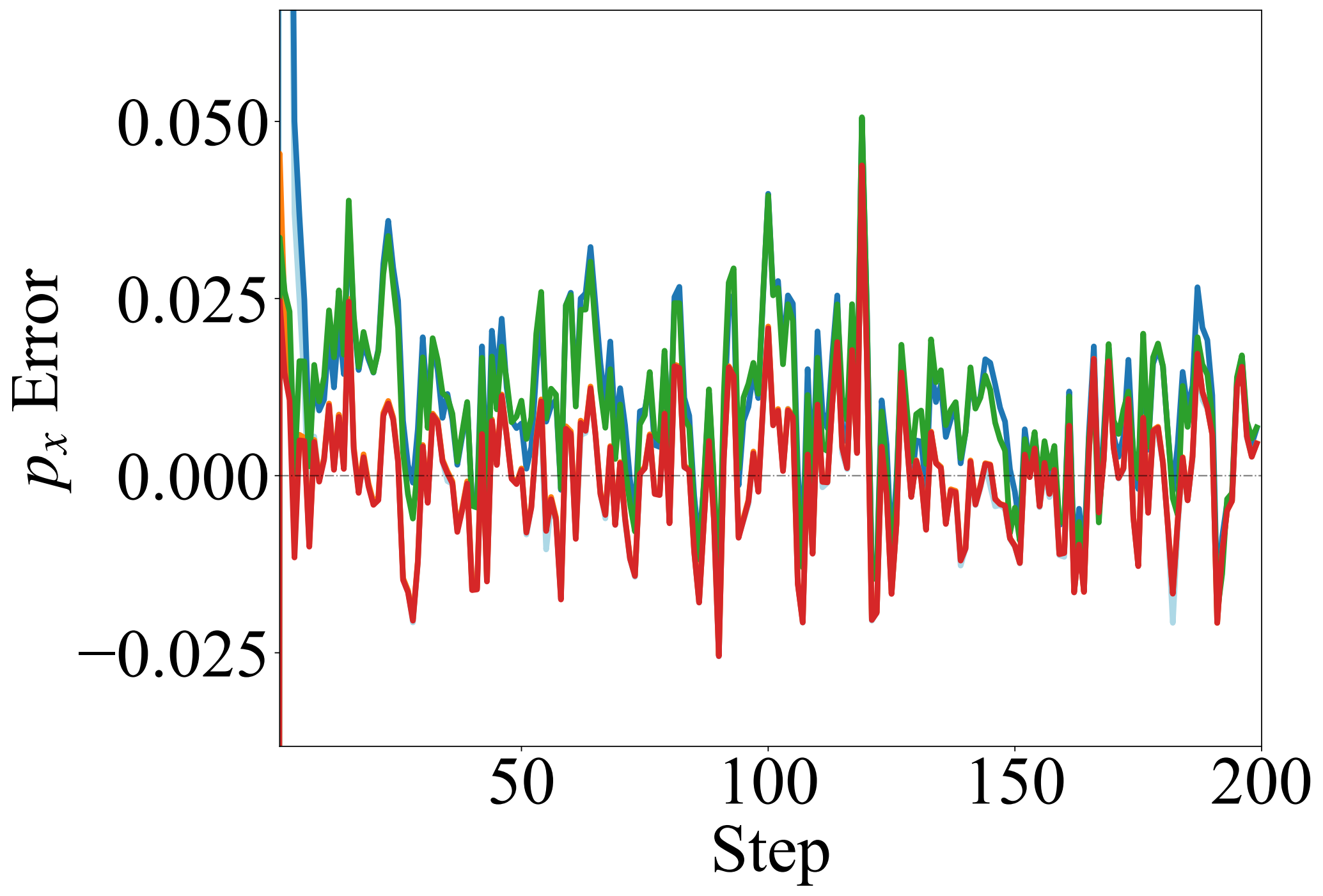}
        \label{localization_la:1}
    \hfill
    \includegraphics[width=0.148\textwidth]{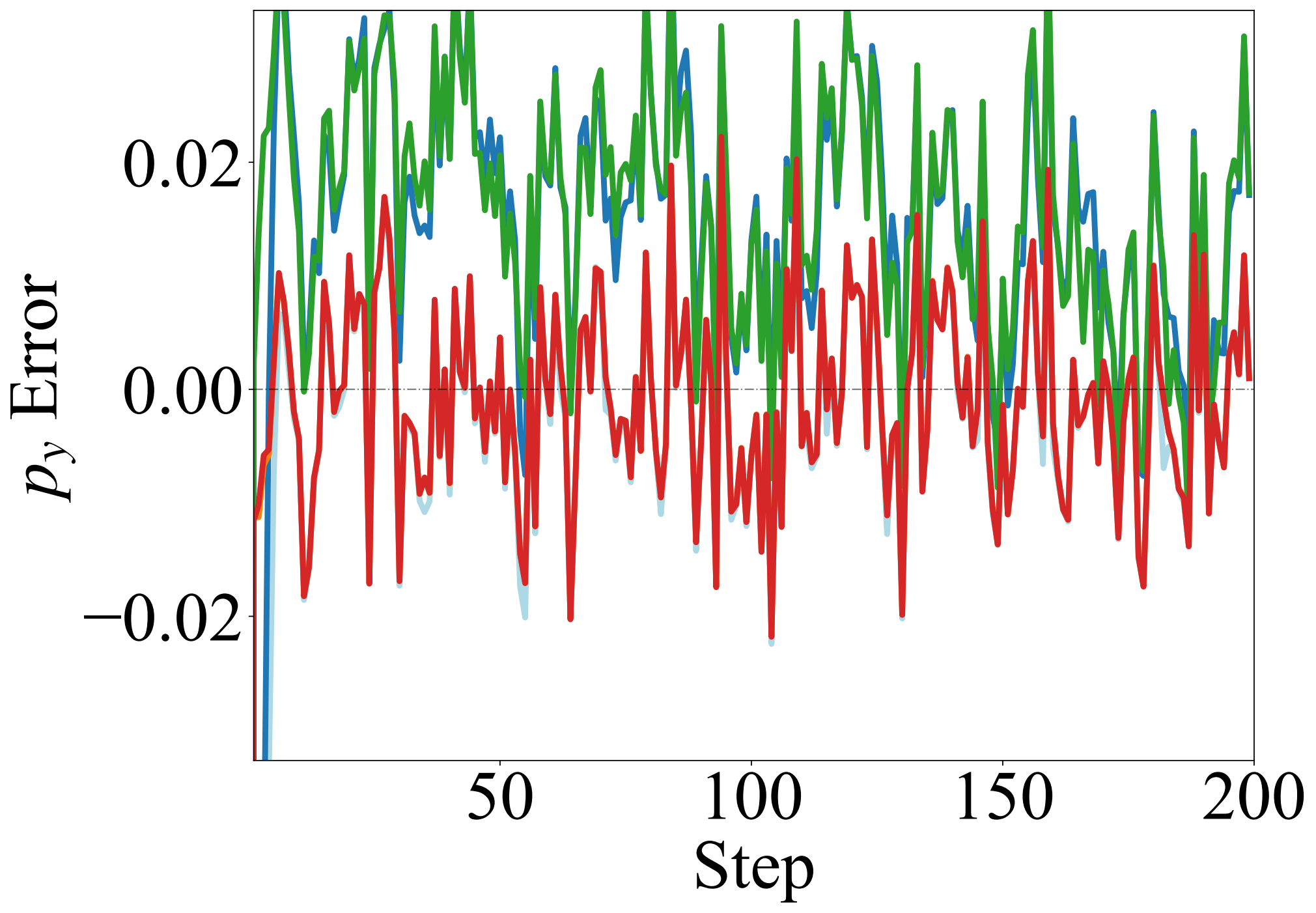}
        \label{localization_la:2}
    \hfill
    \includegraphics[width=0.148\textwidth]{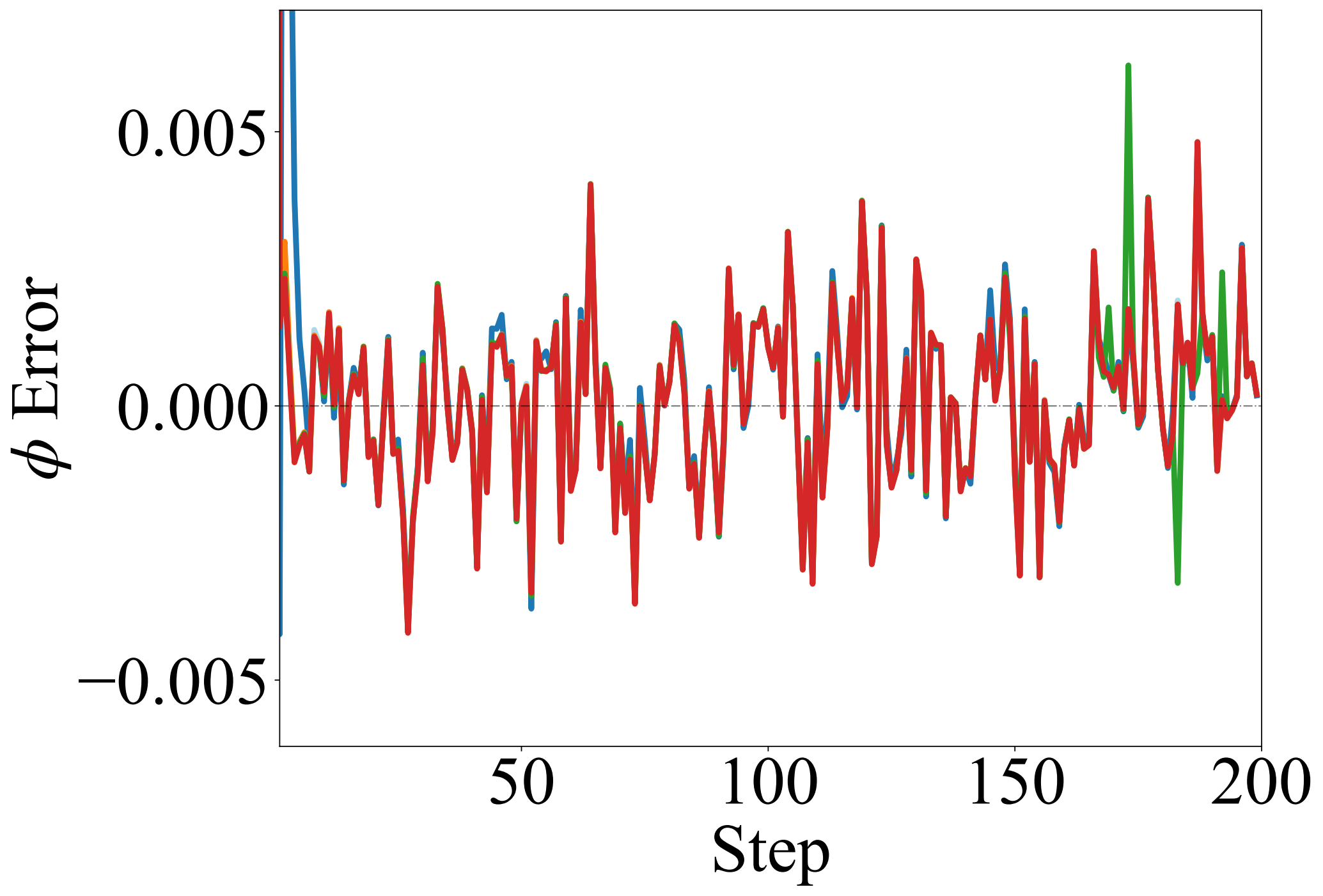}
        \label{localization_la:3}
    }
    \\
    \subfloat[Case C: Beta noise]{
    \includegraphics[width=0.148\textwidth]{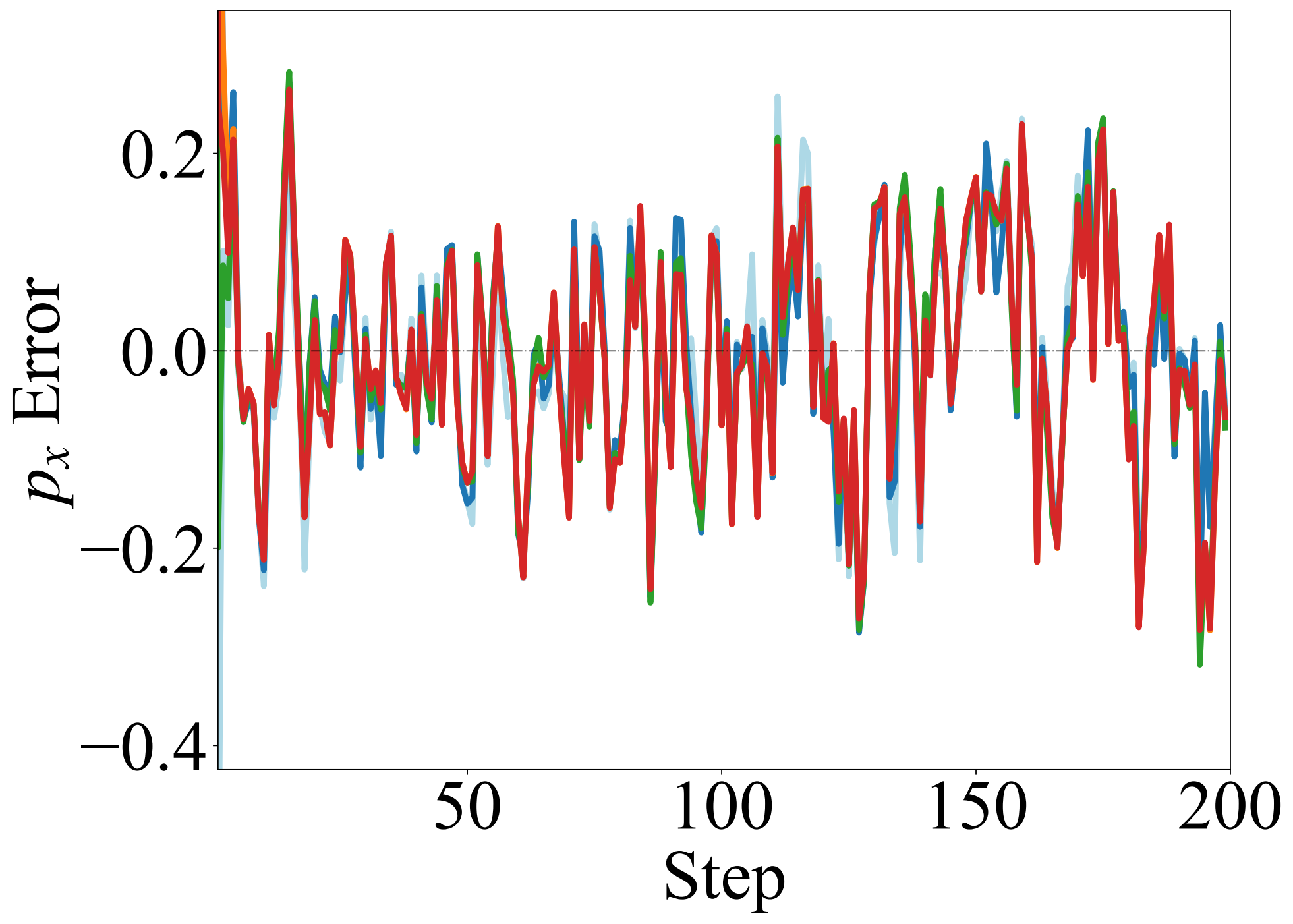}
        \label{localization_beta:1}
    \hfill
    \includegraphics[width=0.148\textwidth]{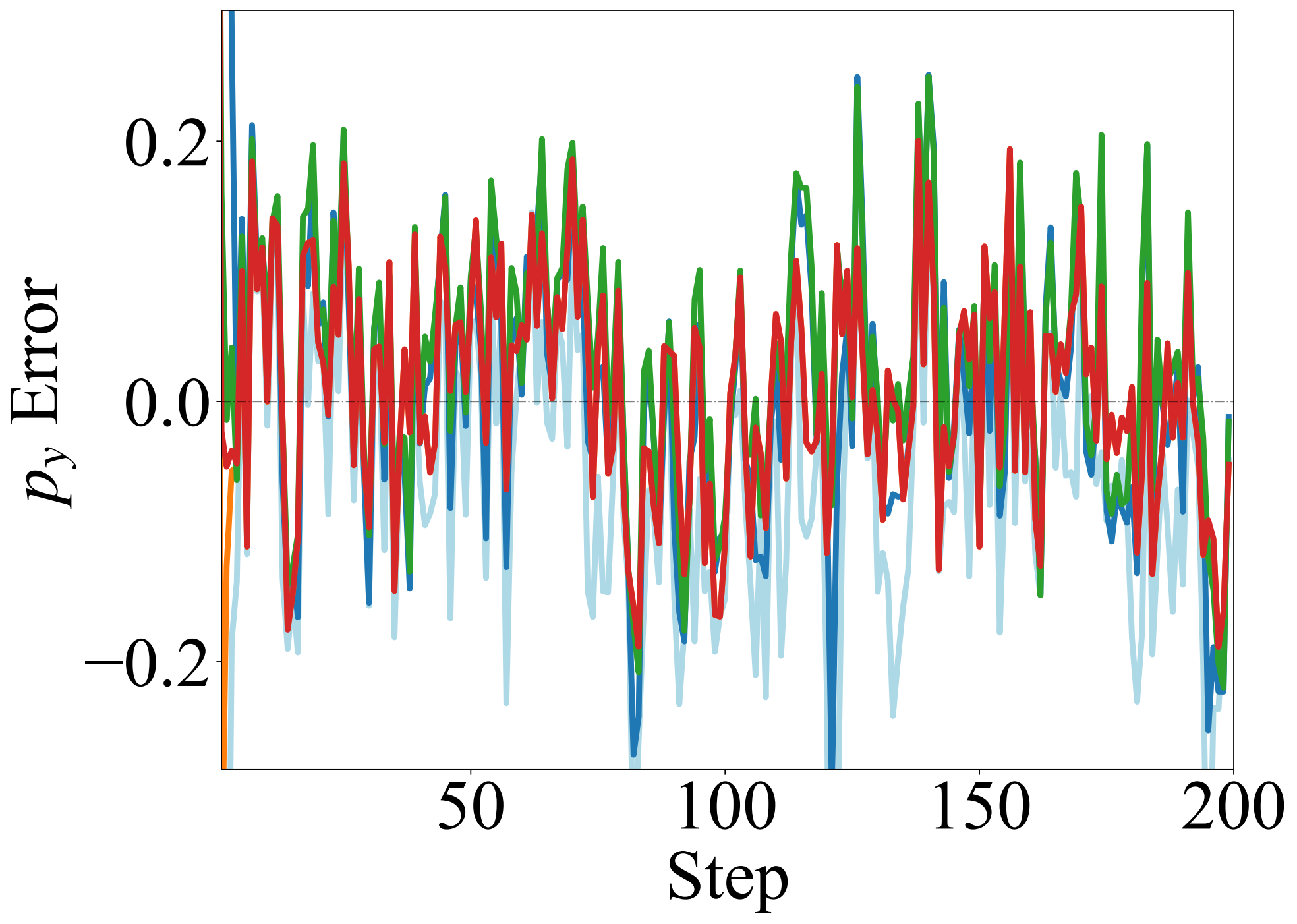}
        \label{localization_beta:2}
    \hfill
    \includegraphics[width=0.148\textwidth]{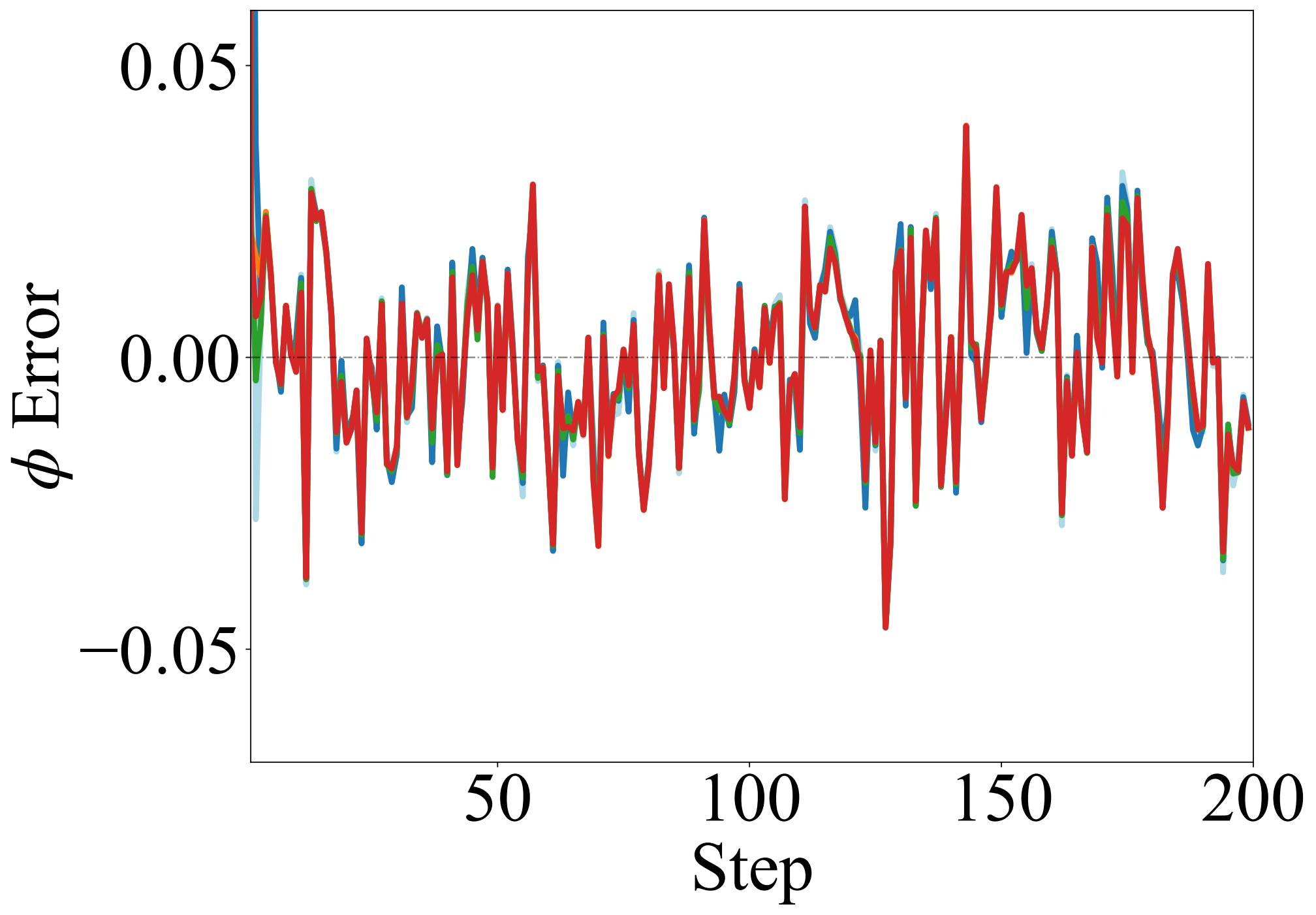}
        \label{localization_beta:3}
    }
    \\
    \subfloat{
        \includegraphics[width=0.45\textwidth]{figures/legend.png}
        \label{localization:legend}
    }
    \caption{Estimation errors in robot localization.}
    \label{fig.localization_error}
\end{figure} 

We consider a robot localization task in industrial applications \citep{barrau2016invariant}. The state vector is the robot pose, i.e., $x = \left[p_x, p_y, \phi\right]^\top$, and the control input is the robot velocity, i.e., $u = \left[v, \omega\right]^\top$. The state space model of this robot is
\begin{equation}\nonumber
\begin{aligned}
\begin{bmatrix}
p_{x,t+1} \\
p_{y,t+1} \\
\phi_{t+1}
\end{bmatrix} &=\begin{bmatrix}
p_{x,t} + v \cos(\phi_{t}) \Delta t \\
p_{y,t} + v \sin(\phi_{t}) \Delta t \\
\phi_{t} + \omega \Delta t
\end{bmatrix} + \xi_t, \\
\boldsymbol{y}_t &= \begin{bmatrix}
R(\phi_t)^\top (p_t - m_1) \\
R(\phi_t)^\top (p_t - m_2) \\
R(\phi_t)^\top (p_t - m_3)
\end{bmatrix}  + \zeta_t.
\end{aligned}
\end{equation}
with
\begin{equation}\nonumber
R(\phi) = \begin{bmatrix}
\cos{\phi} & -\sin{\phi} \\
\sin{\phi} & \cos{\phi}
\end{bmatrix}, \quad p_t = \begin{bmatrix}
p_{x,t} \\
p_{y,t}
\end{bmatrix}, \quad m = \begin{bmatrix}
m_{x} \\
m_{y}
\end{bmatrix}.
\end{equation}
The control input is set to \( v = 5\sin(\frac{\pi}{20} t) \), \( \omega = 3\sin(\frac{\pi}{20} t) \) and  \( \Delta t = 0.1 \). The initial state is set to \( x_0 \sim \mathcal{N}(0, \mathbf{I}_{3 \times 3}) \). Similarly, the process noise $\xi_t$ and measurement noise $\zeta_t$ have three different distributions:
\begin{itemize}
    \item Case A, Gaussian noise: 
    \begin{equation*}
        \xi_t \sim \mathcal{N}(0, 0.01 \cdot\mathbf{I}_{3 \times 3}),  \quad \zeta_t \sim \mathcal{N}(0,  0.01 \cdot\mathbf{I}_{3 \times 3}).
    \end{equation*}
    \item Case B, Laplace noise:
    \begin{equation*}
    \xi_t \sim \text { Laplace }(0,  0.01 \cdot\mathbf{I}_{3 \times 3}),  \quad \zeta_t \sim \text { Laplace }(0,  0.01 \cdot\mathbf{I}_{3 \times 3}).
    \end{equation*}
    \item Case C, Beta noise:
    \begin{equation*}
    \xi_t \sim \text { Beta }(4, 6) \cdot \mathbf{1}_{3\times 1} , \quad 
    \zeta_t \sim \text { Beta }(4, 6) \cdot \mathbf{1}_{3\times 1}.
    \end{equation*}
\end{itemize}

For robot localization tasks, as shown in Fig.~\ref{fig.localization}, the NANO filter achieves the best performance among all filters in terms of average RMSE. Compared to the second-best IEKF, which has a localization average error of $\left[9.71\text{cm}, 8.78\text{cm}\right]$, the NANO filter achieves a localization error of $\left[8.38\text{cm}, 7.72\text{cm}\right]$, representing an improvement of nearly $15\%$ in localization accuracy.

\begin{table*}[!t]
\label{table.run_time}
    \centering
    \caption{Run Time per time step of different algorithms on our experiments (ms)}
    \begin{tabular}{cccccc}
    \toprule
     Algorithms &
     Linear Oscillator &
     Sequence Forecasting &
     Growth Model &
     Robot Localization &
     Satellite Attitude Estimation
     \\
         \midrule
     EKF & 
     0.017 &
     0.159 & 
     0.733 &
     0.241 &
     0.143 \\
     UKF & 
     0.061 &
     0.118 & 
     0.134 &
     0.243  &
     0.202  \\
     iEKF &
     0.034 &
     0.196 & 
     0.759 &
     0.510  &
     0.513  \\
     PLF & 
     0.194 &
     0.251 & 
     0.301 &
     0.530  &
     0.612  \\
     NANO & 
     0.197 &
     0.397 & 
     0.597 &
     2.340  &
     1.750  \\
    \bottomrule
    \end{tabular}

    \label{tab.time_run}
\end{table*}

\subsection{Satellite Attitude Estimation}
We consider a satellite attitude estimation problem where the attitude of the satellite is represented using Euler angles \citep{brossard2020code}. The state variable is defined as \(\boldsymbol{\theta} = \left[\theta_p, \theta_r, \theta_y\right]^\top \in \mathbb{R}^3\), where \(\theta_p, \theta_r, \theta_y\) correspond to the pitch, roll, and yaw angles, respectively. The control input is the angular velocity of the satellite, denoted as \(\omega\in\mathbb{R}^3\). The state space model of this system is 
\begin{equation}\nonumber
\begin{aligned}
\boldsymbol{\theta}_{t+1} &= \boldsymbol{\theta}_{t} + \Omega(\boldsymbol{\theta}_t)\omega_t\Delta t+ \xi_t, \\
y_t &= \begin{bmatrix}
    C(\boldsymbol{\theta}_t)^\top g \\
    C(\boldsymbol{\theta}_t)^\top b 
\end{bmatrix} + \zeta_t,
\end{aligned}
\end{equation}
with \(\Delta t = 0.01s\). The control input is set to  \(\boldsymbol{\omega}_t = \frac{\pi}{18}\sin(2\Delta t\pi t)\cdot \mathbf{1}_{3 \times 1}\), and the vectors \(g=\left[0, 0,-9.81\right]^\top\), \(b=\left[27.75, -3.65, 47.21\right]^\top\) represent the gravitational acceleration and Earth's magnetic field, respectively. The transformation matrices \(\Omega(\theta)\) and \(C(\theta)\) are given by 
\begin{equation}\nonumber
\begin{aligned}
\Omega(\boldsymbol{\theta})  &=\left[\begin{array}{ccc}
1 & \left(s_p s_r\right) / c_p & \left(c_r s_p\right) / c_p \\
0 & c_r & -s_r \\
0 & s_r / c_p & c_r / c_p
\end{array}\right],  \\
C(\boldsymbol{\theta})&=\begin{bmatrix}
c_y c_p & c_y s_p s_r - s_y c_r & c_y s_p c_r + s_y s_r \\
s_y c_p & s_y s_p s_r + c_y c_r & s_y s_p c_r - c_y s_r \\
-s_p & c_p s_r & c_p c_r
\end{bmatrix},
\end{aligned}
\end{equation}
where \(s_i = \sin(\theta_i)\), \( c_i=\cos(\theta_i)\). The initial state is set to \( x_0 \sim \mathcal{N}(0, 10^{-3}\cdot\mathbf{I}_{3 \times 3}) \). In this experiment, we consider a scenario where the process noise $\xi_t$ follows a Laplace distribution with Laplace-distributed outliers of larger variance, and the measurement noise $\zeta_t$ follows Gaussian noise with Beta-distributed outliers, as
\begin{equation*}
\begin{aligned}  
\xi_t &\sim 0.9\text{Laplace}(0, 10^{-5} \cdot\mathbf{I}_{3 \times 3}) + 0.1\text{Laplace}(0, 10^{-2} \cdot\mathbf{I}_{3 \times 3}),  \\
\zeta_t &\sim 0.85\mathcal{N}(0,  10^{-4} \cdot\mathbf{I}_{3 \times 3})+ 0.15\text{Beta}(1.2, 1.5)\cdot \mathbf{1}_{3\times 1}    
\end{aligned}
\end{equation*}
Note that in this setting, where outliers are present in the system, the ground truth in our simulations is generated using the noise model with outliers, $\xi_t$. However, the $Q$ and $R$ matrices for the filters are still configured based on the noise distribution without outliers.

\begin{figure}[!t]
\centering
    \includegraphics[width=0.45\textwidth]{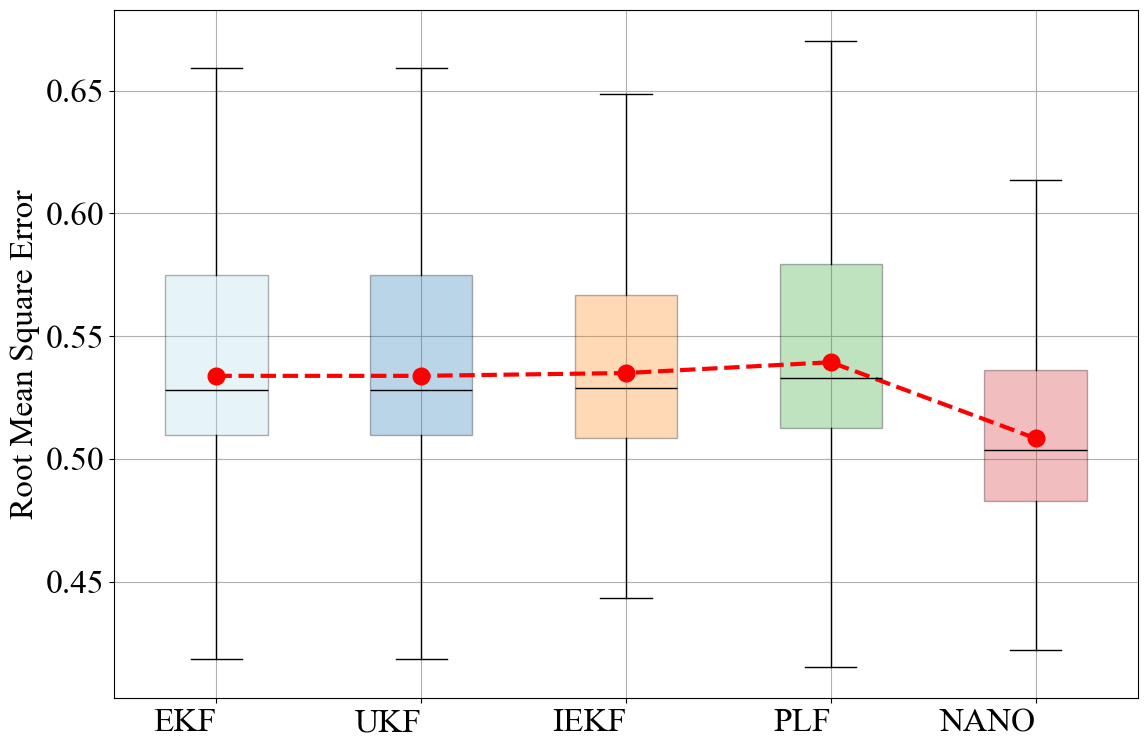}
    \label{fig.Attitude_gauss}

\caption{Box plot of RMSE over all MC experiments for satellite attitude estimation. Note that red point `` $\textcolor{red}{\bullet}$
 " represents the average RMSE.}
\label{fig.attitude}
\end{figure}

\begin{figure}[!t]
    \centering
    \includegraphics[width=0.145\textwidth]{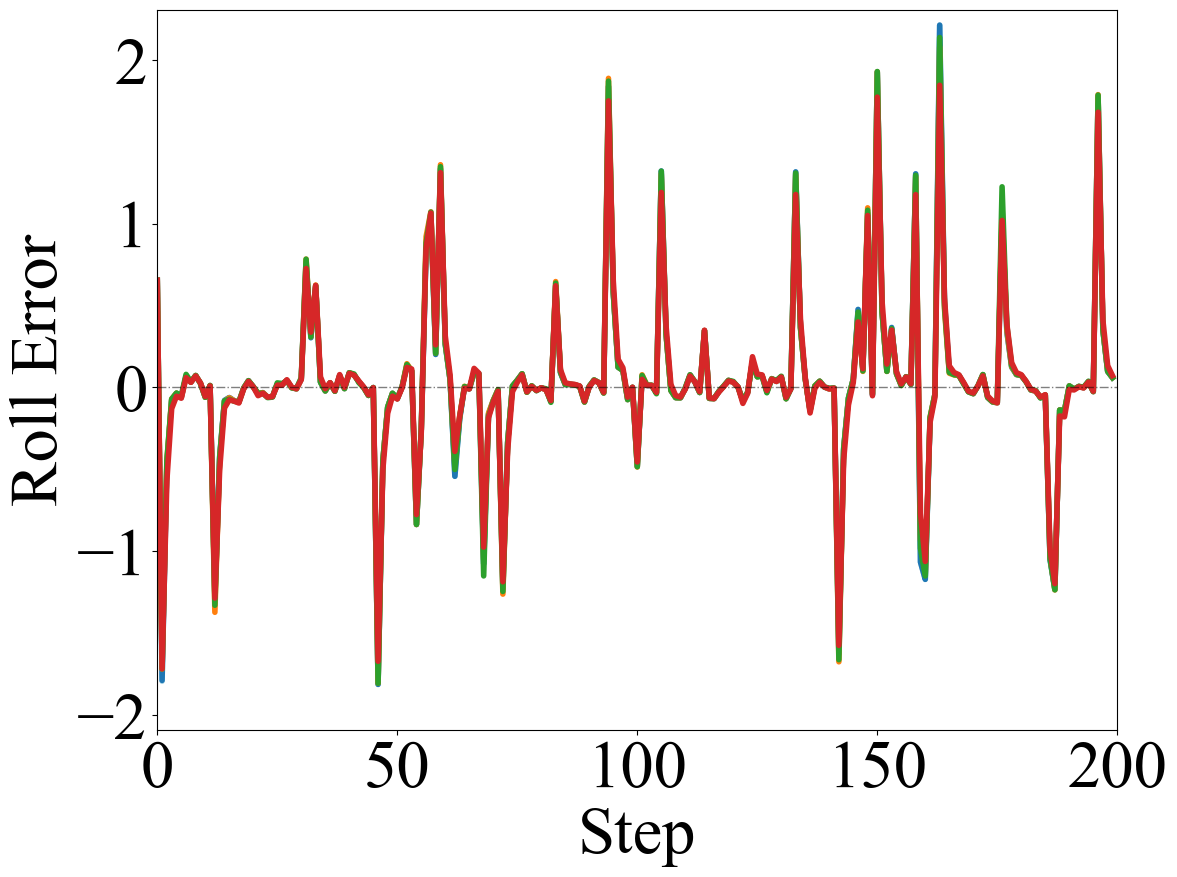}
        \label{Attitude:1}
    \hfill
    \includegraphics[width=0.151\textwidth]{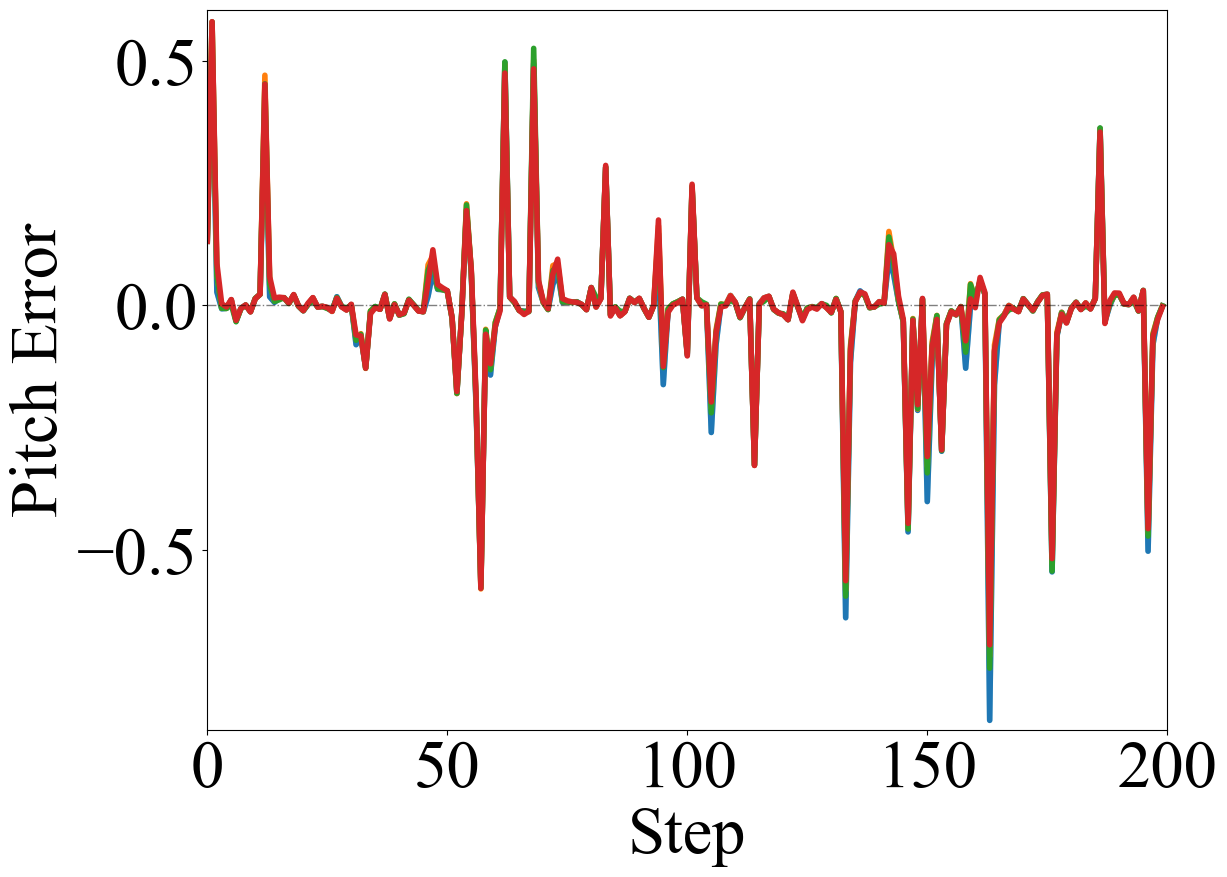}
        \label{Attitude:2}
    \hfill
    \includegraphics[width=0.145\textwidth]{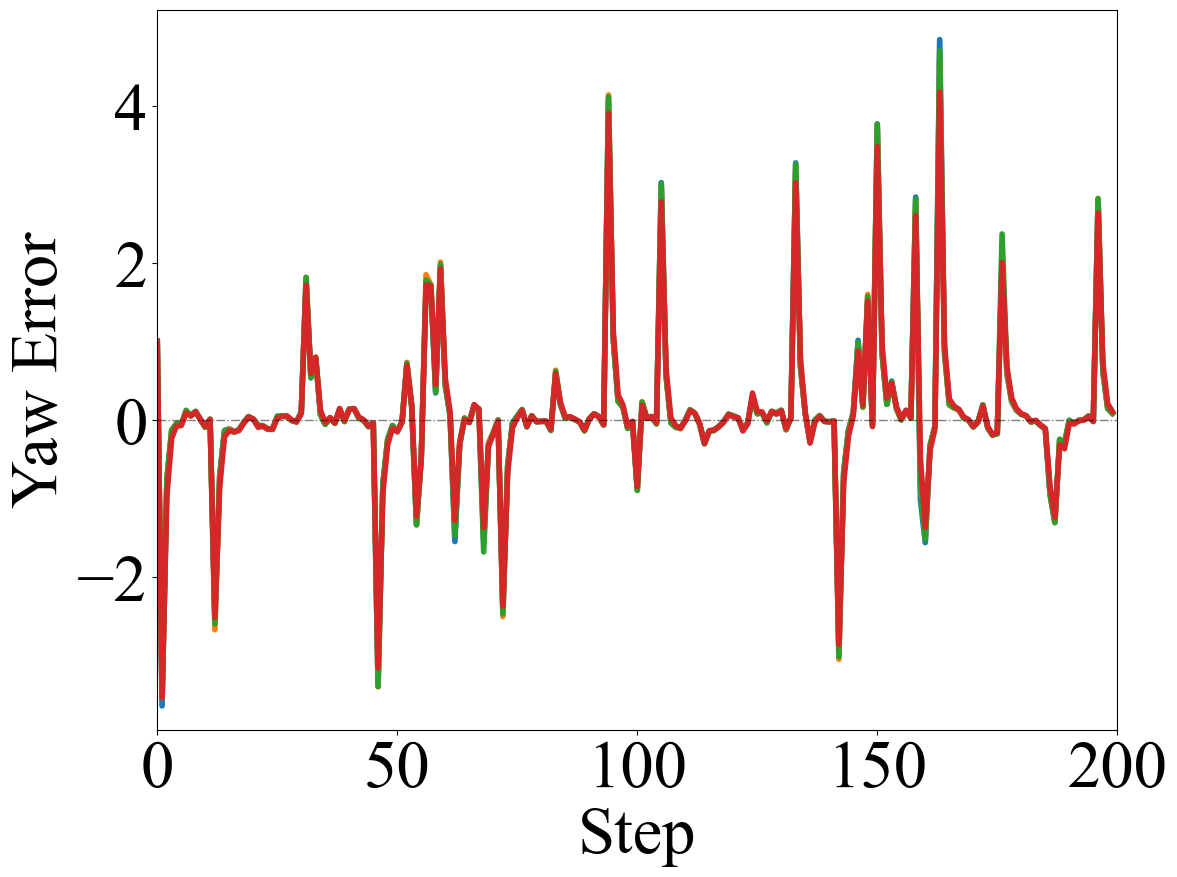}
        \label{Attitude:3}
    \\
    \subfloat{
        \includegraphics[width=0.45\textwidth]{figures/legend.png}
        \label{Attitude:legend}
        
    }
    \caption{Estimation errors in satellite attitude estimation.}
    \label{fig.attitude_error}
\end{figure}

The experimental results are shown in Fig.~\ref{fig.attitude} and Fig.~\ref{fig.attitude_error}. The NANO filter outperforms other methods, demonstrating its stronger robustness compared to general Gaussian filters.

The computational times of five different filters are shown in Table~\ref{tab.time_run}. The NANO filter, being an iterative optimization algorithm, has a higher computational burden compared to the other filters. However, the times are still within a reasonable range, meeting the time requirements for state estimation in practical applications.
\section{Conclusion}
This paper provides an overview of the Natural Gradient Gaussian Approximation (NANO) filter, a new Gaussian filter that improves state estimation in nonlinear systems. By directly minimizing the update cost through natural gradient descent, NANO filter avoids the linearization errors commonly found in traditional Kalman filter families. Comparative tests on various systems show that it achieves the lowest root mean square error compared to the extended Kalman filter, unscented Kalman filter, iterated extended Kalman filter and posterior linearization filter.

\bibliography{ifacconf}

\end{document}